\documentclass[fleqn,usenatbib]{rasti}

\usepackage{newtxtext,newtxmath}
\usepackage[T1]{fontenc}

\DeclareRobustCommand{\VAN}[3]{#2}
\let\VANthebibliography\thebibliography
\def\thebibliography{\DeclareRobustCommand{\VAN}[3]{##3}\VANthebibliography}

\usepackage{graphicx}	% Including figure files
\graphicspath{ {plots/} }
\usepackage{amsmath}	% Advanced maths commands
\usepackage{hyperref}
\usepackage{multirow}
\usepackage{longtable}
\usepackage{xcolor}
\usepackage{newtxtext,newtxmath}

\title[QPE Search With Machine Learning]{Searching for Quasi-Periodic Eruptions using Machine Learning}

\author[R. Webbe et al.]{
Robbie Webbe,$^{1}$\thanks{\url{https://orcid.org/0000-0003-1689-3723}}
and A. J. Young$^{1}$\thanks{\url{https://orcid.org/0000-0003-3626-9151}}
\\
$^{1}$H. H. Wills Physics Laboratory, Tyndall Avenue, Bristol, BS8 1TL, UK
}

\date{Accepted XXX. Received YYY; in original form ZZZ}

\pubyear{2023}

\begin{document}
\label{firstpage}
\pagerange{\pageref{firstpage}--\pageref{lastpage}}
\maketitle

\begin{abstract}
Quasi-Periodic Eruptions (QPEs) are a rare phenomenon in which the X-ray emission from the nuclei of galaxies shows a series of large amplitude flares. Only a handful of QPEs have been observed but the possibility remains that there are as yet undetected sources in archival data. Given the volume of data available a manual search is not feasible, and so we consider an application of machine learning to archival data to determine whether a set of time-domain features can be used to identify further lightcurves containing eruptions. Using a neural network and 14 variability measures we are able to classify lightcurves with accuracies of greater than $94\%$ with simulated data and greater than $98\%$ with observational data on a sample consisting of 12 lightcurves with QPEs and 52 lightcurves without QPEs. An analysis of 83,531 X-ray detections from the XMM Serendipitous Source Catalogue allowed us to recover lightcurves of known QPE sources and examples of several categories of variable stellar objects.
\end{abstract}

% Include between one and six keywords.

%%%%%%%%%%%%%%%%%%%%%%%%%%%%%%%%%%%%%%%%%%%%%%%%%%

%%%%%%%%%%%%%%%%% BODY OF PAPER %%%%%%%%%%%%%%%%%%

\section{Introduction}
\label{sec:intro}

The role of machine learning in astrophysics is becoming progressively important, with the scope and scale of surveys and planned missions resulting in ever-increasing volumes of data and in growing archives for current missions. Citizen science projects, like Galaxy Zoo \citep[e.g.][]{lintott_galaxy_2008} will struggle to cope with the volume of data that is expected to be produced with planned survey missions. There is a pressing need to develop automated tools which can process and reduce volumes of data to manageable amounts. Transient events of many types have been the focus of several machine learning approaches. Due to their fleeting nature and the time sensitive nature of follow-up efforts, automation has the potential to increase the number of transient events detected, and to allow for them to be detected sooner. This will allow a greater proportion of their lifetimes to be monitored and including automation in processing pipelines also allows for such events to be detected before a scientist could interact with the observed data. Approaches in using supervised and unsupervised machine learning to detect and classify a greater proportion of transient events in (near) real time \citep[e.g.][]{narayan_machine-learning-based_2018, muthukrishna_rapid_2019, muthukrishna_dash_2019, muthukrishna_real-time_2022} using optical observations have allowed supernovae and some other classes of transients to be detected during the course of the events, although understandably the accuracy of these techniques increases as more of the events are detected. High energy data presents different challenges, as the statistics underpinning observed data are different due to the typically low count rates. Attempts at detecting X-ray transient sources using supervised learning (Random Forest methods) with the 2XMM and 3XMM Serendipitous Source Catalogues have achieved accuracies of $\simeq 97\%$ \citep{lo_automatic_2014} and  $\simeq 92\%$ \citep{farrell_autoclassification_2015} across multi-class classifications using combinations of time-domain and spectroscopic features. 

With new classes of X-ray transients like Quasi-Periodic Eruptions continuing to be discovered it is important to develop methods for detecting these new classes both in archival data and continuing and planned surveys as soon as possible to develop our understanding of these transients. If it is possible to detect QPEs with established machine learning techniques and variability features used to detect other types of variability, this could significantly increase our known QPE host population.

\subsection{Quasi-Periodic Eruptions}
\label{subsec:intro-qpe}
The first Quasi-Periodic Eruptions (QPEs) were detected in the Active Galactic Nucleus (AGN) GSN 069 by \cite{Miniutti2019}, and four further likely sources were subsequently identified in the extra-galactic sources RX J1301.9+2747 \citep{Giustini2020}, eRASSU J023147.2-102010 (eRO-QPE1) and eRASSU J023448.9-441931 \citep[eRO-QPE2;][]{Arcodia2021}, and XMMSL1 J024916.6-041244 \citep{Chakraborty2021}. Although all AGN show X-ray variability, QPEs are characterised by short lived, large scale changes in X-ray luminosity with eruptions appearing greater in amplitude, peaking at earlier times, and having shorter durations with increasing photon energy bands. Of those objects, two were detected by direct analysis of observations of similar sources \citep[GSN 069 and RX J1301.9+2747;][]{Miniutti2019, Giustini2020}. The sources eRO-QPE1 and eRO-QPE2 were detected by means of a blind search through data released by the eROSITA instrument, with a simple cut by count rate and significant variability being used to identify sources for further examination \citep{Arcodia2021}. The search which identified the QPE candidate source XMMSL1 J024916.6-041244 used the Quasi-periodic Automated Transit Search algorithm \citep{carter_quasiperiodic_2013, Chakraborty2021} which was originally intended for the identification of exoplanet transits. QPEs appear to be a transient phenomena, with the eruptions seen in XMMSL1 J024916.6-041244 not appearing in an observation 15 years after that in which they were observed, and the eruptions in GSN 069 having not been observed in the most recent observations \citep{miniutti_repeating_2022} following an extended period when they were visible.

The mechanism which causes the appearance of QPEs is currently unknown, but possible explanations which have been proposed for the phenomena include: accretion from an orbiting object \citep{King2020,king_quasi-periodic_2022, chen_milli-hertz_2022, krolik_quasiperiodic_2022, lu_quasi-periodic_2022, wang_model_2022, zhao_quasi-periodic_2022}; collision between an orbiting body and the AGN's accretion disc \citep{Xian2021, sukova_stellar_2021}; tearing of warped accretion discs \citep{Raj2021, musoke_disc_2023}; accretion from interacting stellar Extreme Mass Ratio Inspirals (EMRIs) \citep{Metzger2022}; disc instabilities \citep{sniegowska_possible_2020, pan_disk_2022, kaur_magnetically_2022}; gravitational lensing in supermassive black hole binary systems \citep{Ingram2021}. Ultimately, in order to provide a greater evidence base upon which to make rigorous determinations as to the true mechanism for the creation of QPEs we need to identify more sources, and so exploiting future surveys as well as archival data will be important.

\bigskip
In this paper we aim to determine whether lightcurves containing QPEs can be distinguished from those which do not by means of an automated system based upon a series of time-domain variability features. For the purposes of this analysis we will be focusing on the temporal features of Quasi-Periodic Eruptions as seen in the five host objects identified to date. We aim to identify patterns of variability which are quasi-Gaussian in appearance with long periods of quiescence between eruptions. This will allow us to identify future QPE candidates in large survey data or by searching through archival databases. We do this by means of a neural network trained on simulated data and also test the resulting classifier on real observational data from \emph{XMM-Newton}. In section \ref{sec:methods} we describe the generation of the simulated training data sets, the variability features to be used and how the optimal architecture for the classifiers will be determined. We report the results of the classifier against the simulated and real data sets in section \ref{sec:results}. In section \ref{sec:discuss} we discuss the results of the classifier and how it can be used to exploit readily available archival data catalogues, and in section \ref{sec:conc} we consider the performance of the classifier and avenues for future work.

\section{Methods}
\label{sec:methods}

\subsection{Observational Data Preparation}
\label{subsec:data_prep}
\begin{figure}
	\includegraphics[width=\columnwidth]{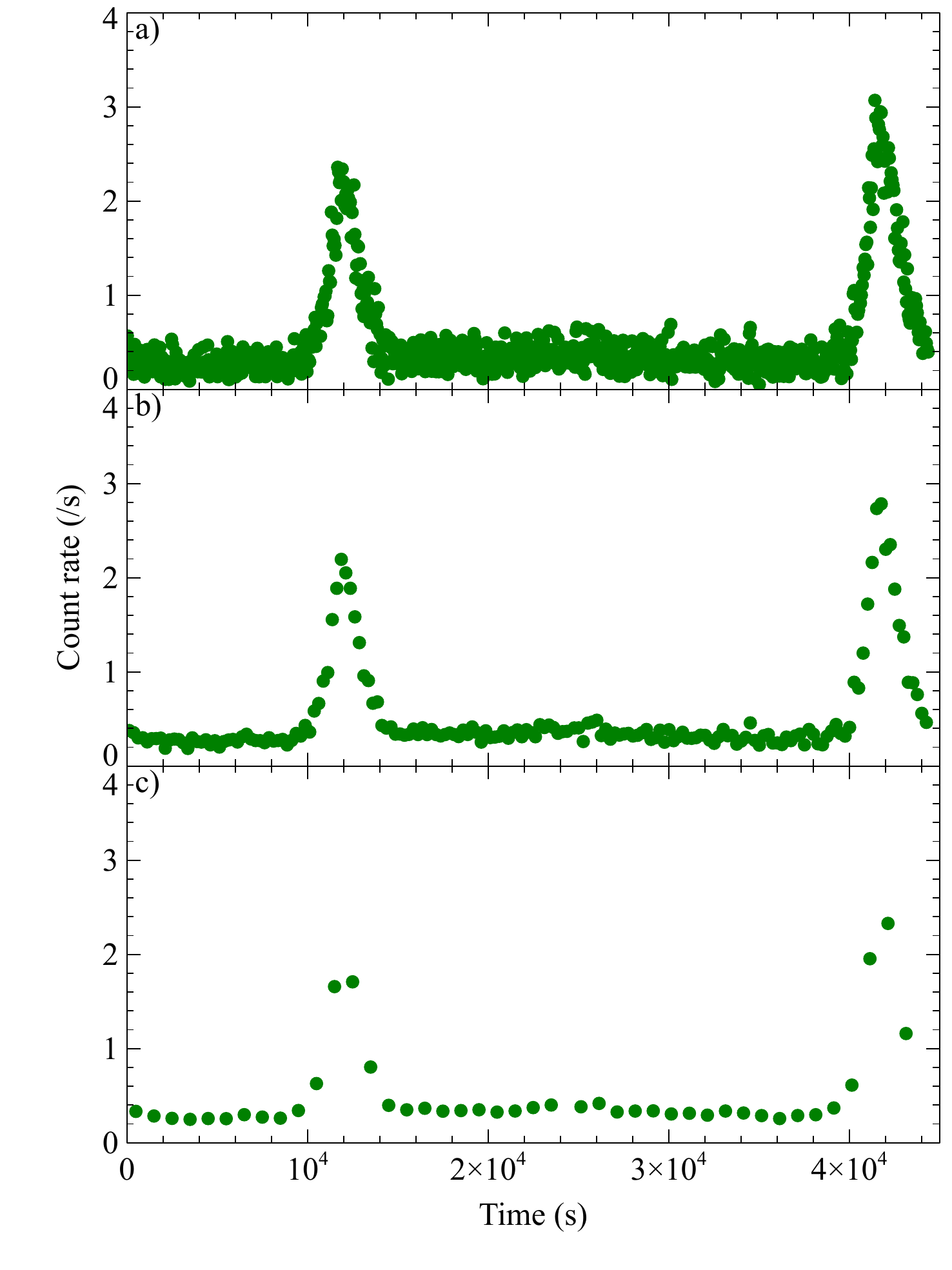}
    \caption{Example of an X-ray lightcurve containing Quasi-Periodic Eruptions (QPEs). Lightcurve is from \emph{XMM-Newton} observation 0823680101 of GSN~069, where the counts are binned at rates of a) 50s, b) 250s, c) 1000s.}
    \label{fig:qpe-ex}
\end{figure}

The training data which we have used in this analysis is derived from a series of \emph{XMM-Newton} observations, listed in Table \ref{tab:qpeml-obs}, of the Quasi-Periodic Eruption sources GSN 069, RX J1301.9+2747, XMMSL1 J024916.6-041244, eRASSU J023147.2-102010, and eRASSU J023448.9-441931. We obtained the data from the XMM Science archive\footnote{\url{http://nxsa.esac.esa.int/nxsa-web/}} and reprocessed all observations using XMM Science Analysis System\footnote{version xmmsas$\_$20190531$\_$1155-18.0.0}. For all lightcurves the EPIC pn camera event data was extracted for photon energies in the range 0.2--2.0 keV, background and barycentre corrected, and was binned at a rate of 10s. We use photon energies in this range as eruptions in this energy range have been easily detectable in known QPE sources and have previously been used to characterise QPE profiles \citep[][etc.]{Miniutti2019,Giustini2020}. The resulting lightcurves were then manually screened for flaring events before being passed for analysis. An example of a screened lightcurve containing QPEs is displayed in Fig. \ref{fig:qpe-ex}.

\begin{table}
	\centering
	\caption{Details of \emph{XMM-Newton} observations of five QPE host galaxies used in generating the training data. Observation exposures and numbers of eruptions contained are listed in the last two columns.}
	\label{tab:qpeml-obs}
	\begin{tabular}{lccr} % five columns, alignment for each
		\hline
		Object & OBSID & Exposure (ks) & Erup.\\
		\hline
		GSN 069 & 0823680101 & 63.3 & 2\\
		-- & 0831790701 & 141.4 & 5\\
            -- & 0851180401 & 135.4 & 5\\
            -- & 0864330101 & 141.0 & 4\\
            RX J1301.9+2747 & 0124710801 & 29.8 & 1\\
            -- & 0851180501 & 48.4 & 3\\
            -- & 0864560101 & 134.9 & 8\\
            XMMSL1 J024916.6-041244 & 0411980401 & 11.7 & 1\\
            eRASSU J023147.2-102010 & 0861910201 & 94.2 & 2\\
            -- & 0861910301 & 90.2 & 1\\
            eRASSU J023448.9-441931 & 0872390101 & 95.0 & 9\\
            -- & 0893810501 & 25.0 & 3\\
		\hline
	\end{tabular}
\end{table}

For each of these observations we then created lightcurve segments which isolated each eruption, giving a total of 43 segments, each containing one eruption, from 12 observations, and fit a model to the segments of the form
\begin{equation}
    x(t)=  x_\text{q} + A \exp \left( \frac{-\ln{(2)}(t-t_\text{peak})^2}{t_\text{dur}^2} \right)
	\label{eq:qpeml-gauss}
\end{equation}
where $x_\text{q}$ is the quiescent baseline countrate in the vicinity of each eruption, $A$ is the amplitude of the eruption relative to this quiescent countrate, $t_\text{peak}$ is the time of the peak of the eruption within the segment, and $t_\text{dur}$ is the duration of the eruption as defined as the FWHM of the Gaussian profile of the eruption. For the purposes of the feature extraction we only include the first eruption seen in observation 0861910201 of eRASSU J023147.2-102010, as the second eruption appears to overlap with the first, and there is no `quiescent' phase on either side of the peak as the observation ends before the eruption has completed. As such there is no baseline against which the peak height could be compared for this second eruption and was therefore not suitable for feature extraction. We accept that excluding the features of this eruption could negatively impact the ability of the classifier to detect other sources with overlapping eruptions. We then used the positions of the peaks within the 8 observations for which there were multiple profiled eruptions to determine the recurrence times between eruptions for each observation, and thus the average duty cycle ($\Delta$) for each observation. The duty cycle is calculated as
\begin{equation}
    \Delta = \frac{t_{\text{dur}}}{t_{\text{rec}}}
	\label{eq:qpeml-dutycycle}
\end{equation}
For the purposes of testing the neural networks against real observational data we also include 57 lightcurves from other low-mass AGN. These lightcurves are of AGN which were selected by \cite{webbe_variability_2023} as part of a targeted search for QPE host galaxies selected by mass from the \emph{Chandra} ACIS Survey for X-Ray AGN in Nearby Galaxies catalogue \citep{She2017} which showed no signs of QPE-like behaviour. Details of all observations used in the final evaluation phase of the neural network performance can be found in Table \ref{tab:all-obs1} in Appendix \ref{appendix:A}.

\subsection{Simulated Lightcurve Generation}
\label{subsec:simlc_gen}
In order to produce a training data set of an appropriate size for the training of the  neural network we used the features of the eruptions already observed (Amplitude, Duration and Duty Cycle) and the algorithm for generating lightcurves outlined in \cite{timmer_generating_1995}. We generate lightcurves with total durations of 100\,ks, and time bins of 50\,s, 250\,s and 1000\,s. We use a value of 250\,s for the time binning as it fits typical binning values used in the Literature, and then the choices of 50\,s and 1000\,s were made as values which could show the variability at the lower level which might be expected to be affected by Poisson noise, and a higher level at which we start to reach the durations of individual eruptive events. With these time bins and considering the typical count rates seen in the data we find that including Poisson noise is not required when training the model (see section~\ref{subsec:further-work}). A simple power-law model was chosen as a typical underlying PSD of the form $f^{-{\beta}}$ with values of $\beta$ being randomly drawn from a normal distribution with $\overline{\beta}=2.06$ and $\sigma_{\beta}=0.01$ as per \cite{Gonzalez-Martin2012}. Values for the index of the power spectra are generated with \texttt{numpy}. As the generated lightcurves have $\bar{x}=0$ we shift the resulting lightcurves up by $(10+\delta)x_{\text{min}}$, where $\delta$ is sampled from a Normal distribution, and $x_{\text{min}}$ is the lowest point in the raw lightcurve. In order to create a sub-population of lightcurves which contain quasi-periodic eruption signals we then multiply half of the simulated lightcurves with a signal of the form
\begin{equation}
    x(t) = 1 + A \sum_{m=1}^{M} \exp \left( \frac{-\ln{(2)}(t-t_0-[m-1]t_\text{rec})^2}{t_{\text{dur}}^2} \right).
	\label{eq:qpeml-qpesignal}
\end{equation}
where $A$ is the amplitude of the eruptions being modelled, $t_{\text{dur}}$ is the duration, $t_0$ is the peak time of the first eruption in the series, $t_{\text{rec}}$ is the recurrence time between eruption peaks, and where $M$ satisfies
\begin{equation}
        M = \left\lfloor \frac{T}{t_{\text{rec}}} \right\rfloor + 3
\end{equation}
This gives us a balanced data set for training and validation. To simulate a range of possible eruption profiles we randomly sample values for $A$, $t_0$, $t_\text{dur}$, and $\Delta$ on the basis of the eruptions profiled in section \ref{subsec:data_prep}. Values for $t_0$ are sampled from a uniform distribution such that $t_0 \in[-t_{\text{rec}},0)$, and the values for $A$, $t_\text{dur}$ and $\Delta$ are sampled from exponentially modified Gaussian distributions \citep[e.g.][]{gladney_computer-assisted_1969,grushka_characterization_1972}. We sample from such a distribution as it provided a high quality fit to the distribution of observed parameter values. These simulated lightcurves are then processed to create non-parametric statistical features, as outlined in section \ref{subsec:var_params}.

\subsection{Variability Measures}
\label{subsec:var_params}
We will be using a set of variability statistics calculated from our simulated sample, and then from the real testing sample, to determine whether a lightcurve does or does not contain quasi-periodic eruptions. Some of these features have been used previously in attempts to characterise lightcurve variability \citep{sokolovsky_comparative_2017}, and some have been used in coordination with machine learning techniques like self-organising maps \citep{faisst_how_2019} to characterise AGN variability. Due to the use of simulated lightcurves in the training and validation steps of our machine learning workflow we do not have errors on the simulated lightcurves. As such, we choose 14 features which do not rely on the existence of uncertainties within the data being analysed. We extract these features from the simulated lightcurves and the observational data at all three time bins of 50\,s, 250\,s and 1000\,s in order to observe the effects of time binning on detection accuracy.

\subsubsection{Feature 1 - Standard Deviation normalised by the Mean}
In order to remove any issue caused by the different count rates for observations we use the standard deviation divided by the mean count rate
\begin{equation}
        f_1 = \frac{\sqrt{\frac{\sum_i (x_i-\bar{x})^2}{N-1}}}{\bar{x}}
\end{equation}
where $N$ is the number of points in the lightcurve, $x_i$ are the count rates of individual points on the lightvcurve, and $\bar{x}$ is the average count rate.

\subsubsection{Features 2--7 - Proportion of the lightcurve further than 1$\sigma$ to 6$\sigma$ from the mean}
A lightcurve where the points are normally distributed about a mean count rate will have proportions of points at given numbers of $\sigma$ from the mean count rate. These features are calculated as
\begin{equation}
        f_n = \frac{N_{x,(n-1)\sigma}}{N} , 
        n \in [2,7]
\end{equation}
where $N_{x,(n-1)\sigma}$ is the number of points where $|x_i - \bar{x}| > (n-1)\sigma$.

\subsubsection{Feature 8 - Inter-Quartile Range normalised by $\sigma$}
For a lightcurve where the points are normally distributed about the mean the middle 50\% of points should be within 0.674$\sigma$ of the mean, and so it should be that $IQR = 0.674\sigma$.
\begin{equation}
        f_8 = \frac{IQR}{\sigma}
\end{equation}

\subsubsection{Feature 9 - Skew}
The skewness of a lightcurve will determine how assymetrically the individual count rates are around the mean value. For this analysis we use the Fisher-Pearson coefficient of skewness
\begin{equation}
        f_9 = g_1 = \frac{m_3}{m_2^{3/2}}
\end{equation}
where
\begin{equation}
\label{eq:stat-moment}
        m_i = \frac{1}{N}\sum(x-\bar{x})^i
\end{equation}
and $m_i$ is the bias sampled $i^{\text{th}}$ central moment.

\subsubsection{Feature 10 - Kurtosis}
The kurtosis of a lightcurve will determine, relatively, how likely it is that the lightcurve will contain extreme outlying values. For this analysis we use the Fisher coefficient of kurtosis
\begin{equation}
        f_{10} = g_2 = \frac{m_4}{m_2^2} - 3
\end{equation}
where $m_4$ and $m_2$ are calculated as described in equation Eq. \ref{eq:stat-moment}.

\subsubsection{Feature 11 - Reverse Cross-Correlation normalised by $\sigma$}
To provide a second measure of the assymmetry of the lightcurves being analysed we calculate the cross-correlation for the observations with themselves, being reversed along the time axis. To mitigate for different average count rates across observations we normalise all deviations for individual data points by the standard deviation for the observations as a whole.
\begin{equation}
        f_{11} = \sum_i\frac{(x_i-\bar{x})(x_{N-i}-\bar{x})}{\sigma^2}
\end{equation}
where $x_i$ are the count rates of individual points in the lightcurve, $x_{N-i}$ is the counterpart point on the reversed lightcurve, and $N$ is the number of points in the lightcurve.

\subsubsection{Feature 12 - First maximum of the Autocorrelation Function}
As a measure of coherent periodic variability we take the height of the first peak of the normalised autocorrelation function. This is the first peak after the autocorrelation function has crossed zero, and is normalised by the autocorrelation at $t=0$.
\begin{equation}
        f_{12} = \max\left(\frac{ACF(t)}{ACF(t=0)}\right)
\end{equation}

\subsubsection{Feature 13 - Consecutive Same Sign Deviation (CSSD) proportion}
We consider the proportion of sets of three consecutive points from the lightcurve which all display the same sign deviation from the mean count rate. In some examples in the literature \citep[e.g.][]{wozniak_difference_2000, shin_detecting_2009} a choice is made to require all three points to be at least some multiple of $\sigma$ from the mean value, but we choose not to include this extra distinction in order to not inadvertently prejudice our results, or to cause degeneracies with Features 2-7.
\begin{equation}
        f_{13} = \frac{N_{3x}}{N-2}
\end{equation}
where $N_{3x}$ is the number of sets of three consecutive points in the lightcurve which have the same sign deviation from $\bar{x}$.

\subsubsection{Feature 14 - Von Neumann ratio}
The final feature compares the difference in the deviation between successive pairs of points to the total variance of the lightcurve.
\begin{equation}
        f_{14} = \eta = \frac{\sum_{i=1}^{N-1}(x_{i+1}-x_i)^2/(N-1)}{\sigma^2}
\end{equation}
where $x_i$ and $x_{i+1}$ are pairs of consecutive points.

\subsection{Neural Network Architecture}
\label{subsec:nn-arch}
The neural networks being applied and evaluated in this work are built using the \texttt{tensorflow} \citep{tensorflow2015-whitepaper} package for \texttt{python}. We create networks with an input layer, either one or two hidden layers and an output layer. The input has nodes as required by the number of input features, and the output layer has two nodes to force a binary classification between lightcurves containing QPEs or not. We use the \texttt{relu} activation function, and as we are considering a classification problem our loss function of choice is \texttt{Cross Entropy}. Outputs from the network are then passed to a \texttt{SoftMax} layer to give the probability that lightcurves belong to the classes. The hidden layers are allowed to have between 3 and 196 nodes, and the precise final architecture for each input data set is determined by \texttt{keras\_tuner} \citep{omalley2019kerastuner}. We use the \texttt{Hyperband} tuner from \texttt{keras\_tuner} for this optimisation. The training and evaluation of networks with the simulated data uses accuracy as the metric to be optimised, which forms part of the determination as to the optimal architecture.
\begin{equation}
        \text{Accuracy} = \frac{TP + TN}{TP + FP + TN + FN}
\end{equation}
where $TP$ is the number of true positive, $FP$ is the number of false positive, $TN$ is the number of true negative, and $FN$ is the number of false negative classifications. For these purposes we consider those lightcurves which contain QPEs as being classified as positive, with a probability, as determined by the neural network, of containing QPEs being greater than 50$\%$. When we evaluate the performance against the real data sets we also consider purity, completeness and the $F_{\text{1}}$ score for each network where 
\begin{equation}
        \text{Purity} = \frac{TP}{TP+FP} ,
\end{equation}
\begin{equation}
        \text{Completeness} = \frac{TP}{TP+FN} ,
\end{equation}
and
\begin{equation}
        \text{$F_{\text{1}}$ Score} = \frac{2 \times \text{Purity} \times \text{Completeness}}{\text{Purity} + \text{Completeness}}
\end{equation}
We use these additional measures as the observational data set is less balanced, with only 17.4$\%$ of the real observations containing QPEs. These metrics are not used to inform further amendments to the architecture of the network, or the weightings applied to individual nodes, but are simply used to illustrate the effectiveness of the network in classifying real observational data.

\section{Results}
\label{sec:results}

\subsection{Simulated Lightcurve Classification}
\label{subsec:simlcres}
We simulated 100,000 lightcurves, with 50,000 to contain QPEs, and 50,000 without QPEs, with a duration of 100\,ks and time binning of 50\,s, 250\,s and 1\,ks in the manner described in section \ref{subsec:simlc_gen} and extracted from them the variability features described in section \ref{subsec:var_params}. From the population of 100,000 simulated lightcurves at each time binning rate we set aside 10,000 lightcurves for testing and split the remainder in the ratio 80:20 for training and validation.

The optimal architectures for the neural networks were determined as described in section \ref{subsec:nn-arch} and allowed to vary between different values for the time binning both for the number of nodes in any hidden layers, and by the number of hidden layers which were allowed to exist. The optimal architectures for the two lower time binned networks included only one hidden layer, while the 1ks binned network contained two hidden layers. In all cases there was an output layer which was fixed to contain 2 nodes in order to force a binary output choice between the lightcurve containing QPEs or not. The full architectures are given in Table \ref{tab:nn-arch}.
\begin{table}
	\centering
	\caption{Optimal architecture for the neural networks at each time binning value as determined by \texttt{keras$\_$tuner}.}
	\label{tab:nn-arch}
	\begin{tabular}{lcr} % five columns, alignment for each
		\hline
		Time Bin (s) & No. Hidden Layers & Nodes \\
		\hline
		50 & 1 & [110, 2] \\
		250 & 1 & [49, 2] \\
            1000 & 2 & [23, 91, 2] \\
		\hline
	\end{tabular}
\end{table}

Across all three time bin values we achieved very high levels of accuracy when classifying the simulated data sets, with the validation and testing accuracy being greater than 94$\%$ in all cases and are given in Table \ref{tab:simlc-acc}.
\begin{table}
	\centering
	\caption{Classification accuracy for 18,000 simulated lightcurves in the validation phase, and 10,000 in the testing phase.}
	\label{tab:simlc-acc}
	\begin{tabular}{lcr} % five columns, alignment for each
		\hline
		Time Bin (s) & Validation & Testing \\
		\hline
		50 & 0.9404 & 0.9379 \\
		250 & 0.9489 & 0.9502 \\
            1000 & 0.9467 & 0.9495 \\
		\hline
	\end{tabular}
\end{table}
Validation and testing accuracy were greatest for the 250s binned data sets. Differences in performance are, however, only marginal and the slight changes in accuracy upon classifying the simulated testing data set suggests that the model was not significantly overfit to the training data.

\subsection{Observational Data Classification}
\label{subsec:reallcres}

The observational data, as listed in Table \ref{tab:all-obs1}, were rebinned to rates of 50s, 250s and 1000s and features were extracted for all available lightcurves. There were three observations without QPEs which could not be rebinned at a rate of 250s, and a further two which could not be rebinned at a rate of 1000s as well due to the limitations on good time intervals.

\begin{table*}
	\centering
	\caption{Classification accuracy for observational XMM lightcurves of objects as listed in Table \ref{tab:all-obs1}. Classifications are made on the basis of a cut at 50$\%$ threshold probability for containing a QPE or not.}
	\label{tab:simlc_acc}
	\begin{tabular}{lcccccr} 
		\hline
		Time Bin (s) & No. with QPE & No. without QPE & Accuracy & Purity & Completeness & $F_{\text{1}}$ Score \\
		\hline
		50 & 12 & 57 & 0.319 & 0.203 & 1.0 & 0.338 \\
		250 & 12 & 54 & 0.212 & 0.188 & 1.0 & 0.316 \\
            1000 & 12 & 52 & 0.328 & 0.208 & 0.917 & 0.338 \\
		\hline
	\end{tabular}
\end{table*}

\begin{figure}
	\includegraphics[width=\columnwidth]{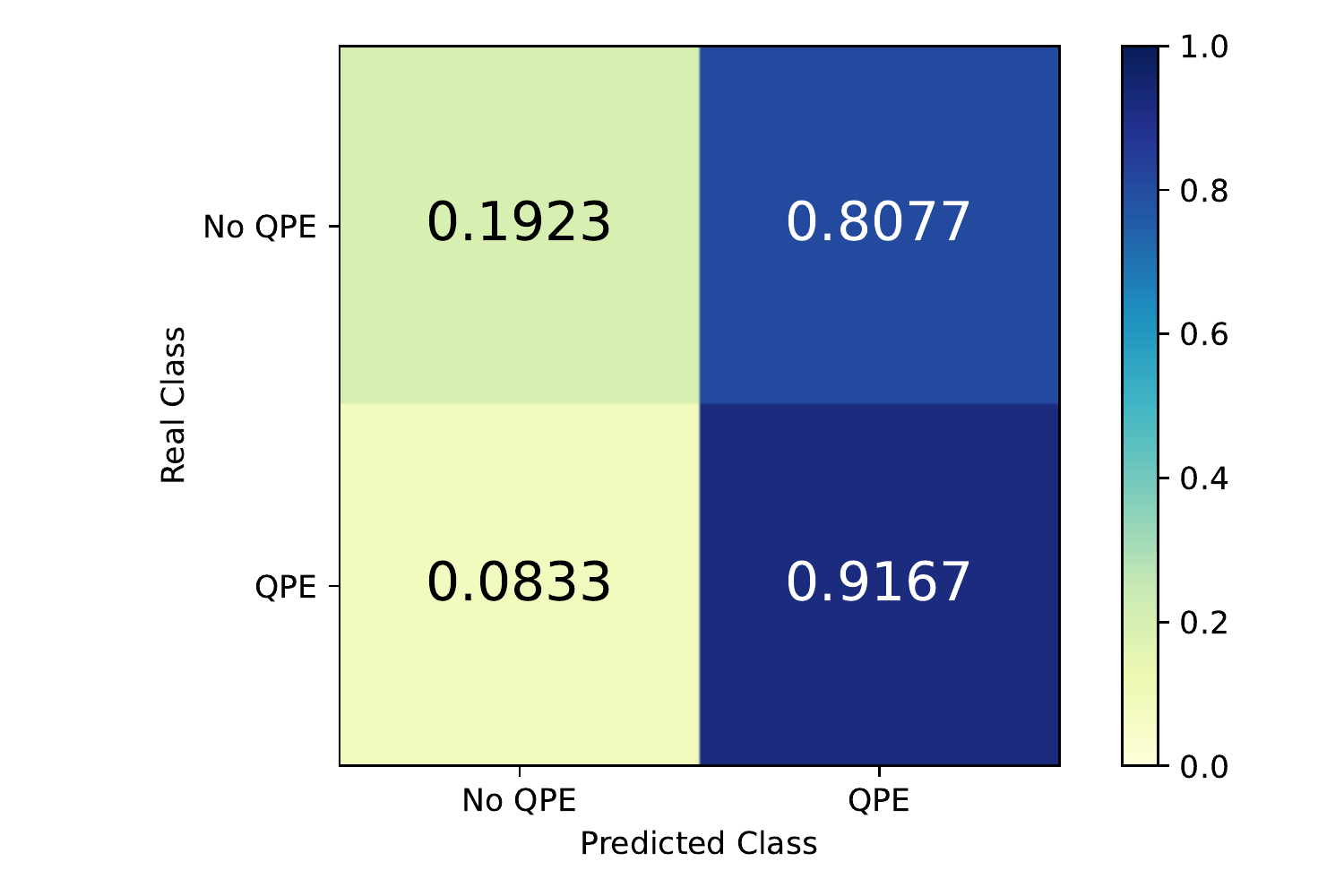}
    \caption{Confusion matrix showing the classification of 52 lightcurves without QPEs and 12 with lightcurves with QPEs present. The lightcurves were binned at a rate of 1000s, and their full details are given in \ref{tab:all-obs1}. Threshold probability for requiring a QPE is set at 0.500.}
    \label{fig:dt1000-confmat}
\end{figure}

There is no clear hierarchy when classifying real observational data, and at first it appears that all three sets of observational data are classed with equally poor performance, although the 250s binning performs slightly below the other two classifiers. The completeness of close to 1.0 and low levels of purity suggest that all lightcurves are being strongly predicted as containing QPEs.
\begin{figure}
	\includegraphics[width=\columnwidth]{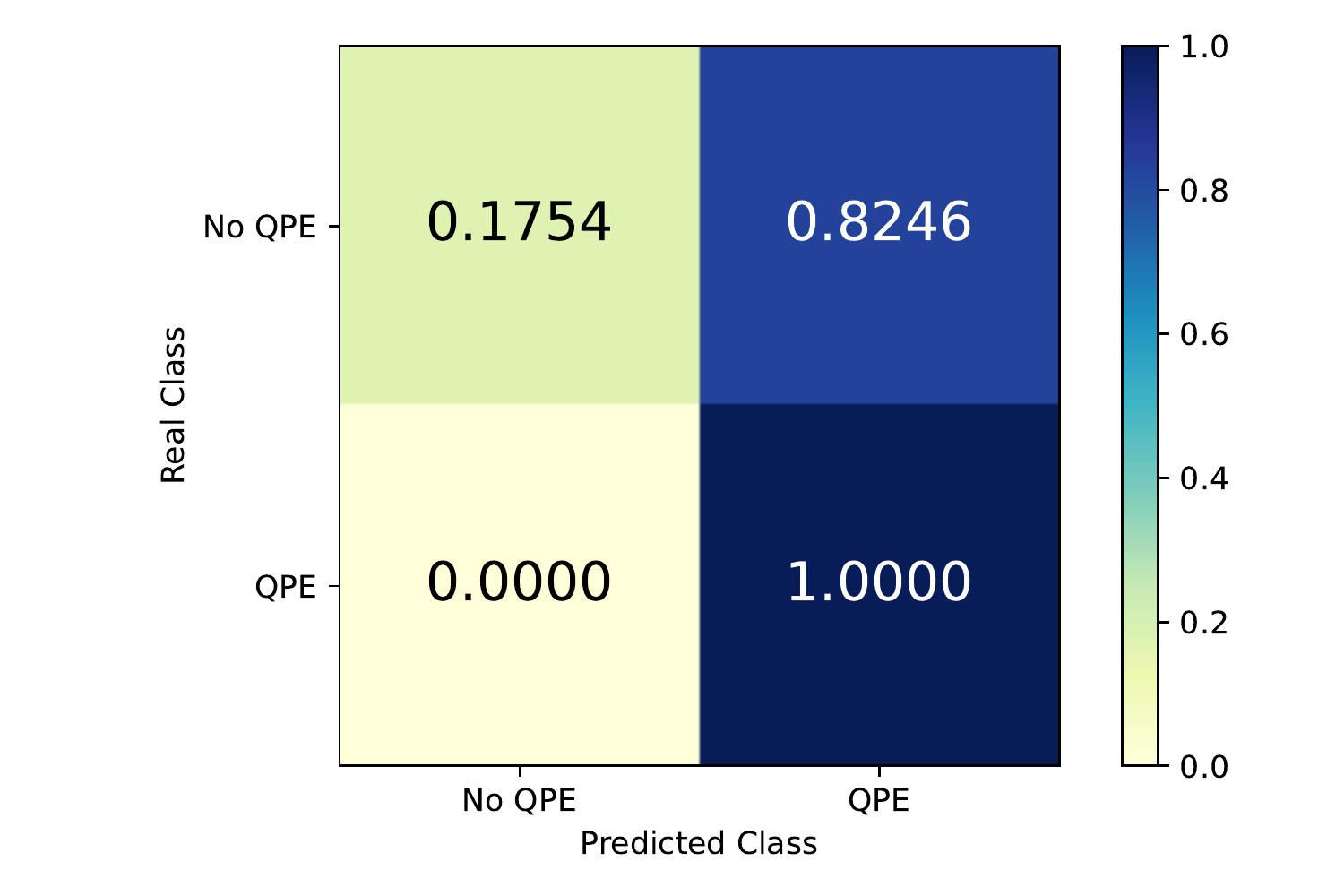}
    \caption{Confusion matrix showing the classification of 57 lightcurves without QPEs and 12 with lightcurves with QPEs present. The lightcurves were binned at a rate of 50s, and their full details are given in \ref{tab:all-obs1}. Threshold probability for requiring a QPE is set at 0.500.}
    \label{fig:dt50-confmat}
\end{figure}
\begin{figure}
	\includegraphics[width=\columnwidth]{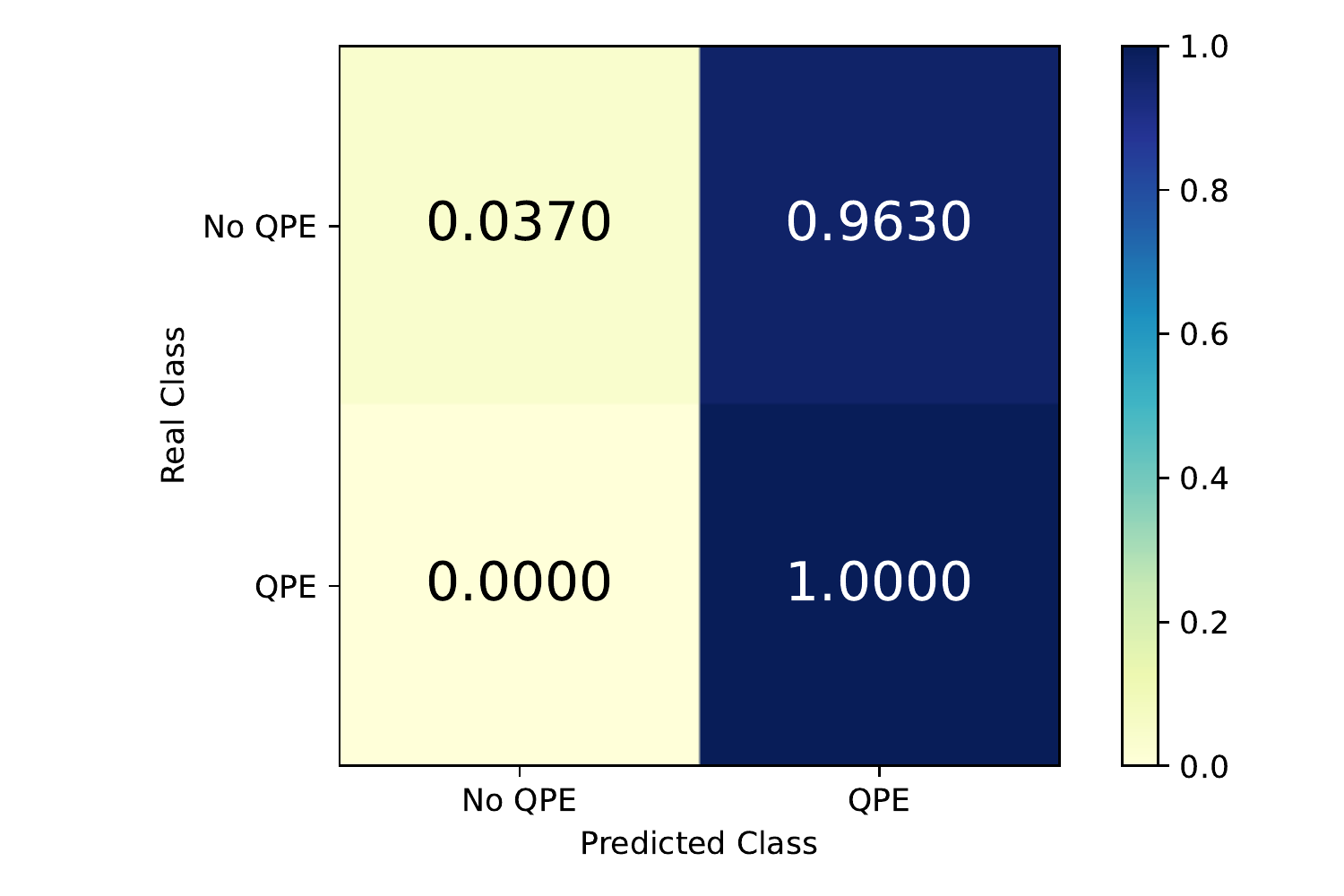}
    \caption{Confusion matrix showing the classification of 54 lightcurves without QPEs and 12 with lightcurves with QPEs present. The lightcurves were binned at a rate of 250s, and their full details are given in \ref{tab:all-obs1}. Threshold probability for requiring a QPE is set at 0.500.}
    \label{fig:dt250-confmat}
\end{figure}
The drop in accuracy when classifying real observational data rather than simulated data suggests at first that there may be features of the real observational data which are not appropriately captured in the simulated data sets. The confusion matrices showing the classifications of the observational data against all three classifiers are show in Fig. \ref{fig:dt50-confmat}, \ref{fig:dt250-confmat}, and \ref{fig:dt1000-confmat}.
From the confusion matrices we can see that the overall accuracy is lowered by large numbers of lightcurves being incorrectly classified as containing QPEs. When we plot the QPE probabilities of the three classifiers we can see that the QPE probabilities generally are quite high and close to 1.0 in the majority of cases, but that those lightcurves containing QPEs do tend to be classed more strongly as containing QPEs than those without, and as such we should consider the level at which we delineate between QPE and non-QPE containing lightcurves. We do consider the benefits of probability calibration, as outlined in \cite{guo_calibration_2017}, but find that calibration by means of isotonic regression using \texttt{sklearn} \citep{scikit-learn} does not beneficially redistribute the probabilities, with the granulation of probabilities close to 1.0 being lost.
\begin{figure}
	\includegraphics[width=\columnwidth]{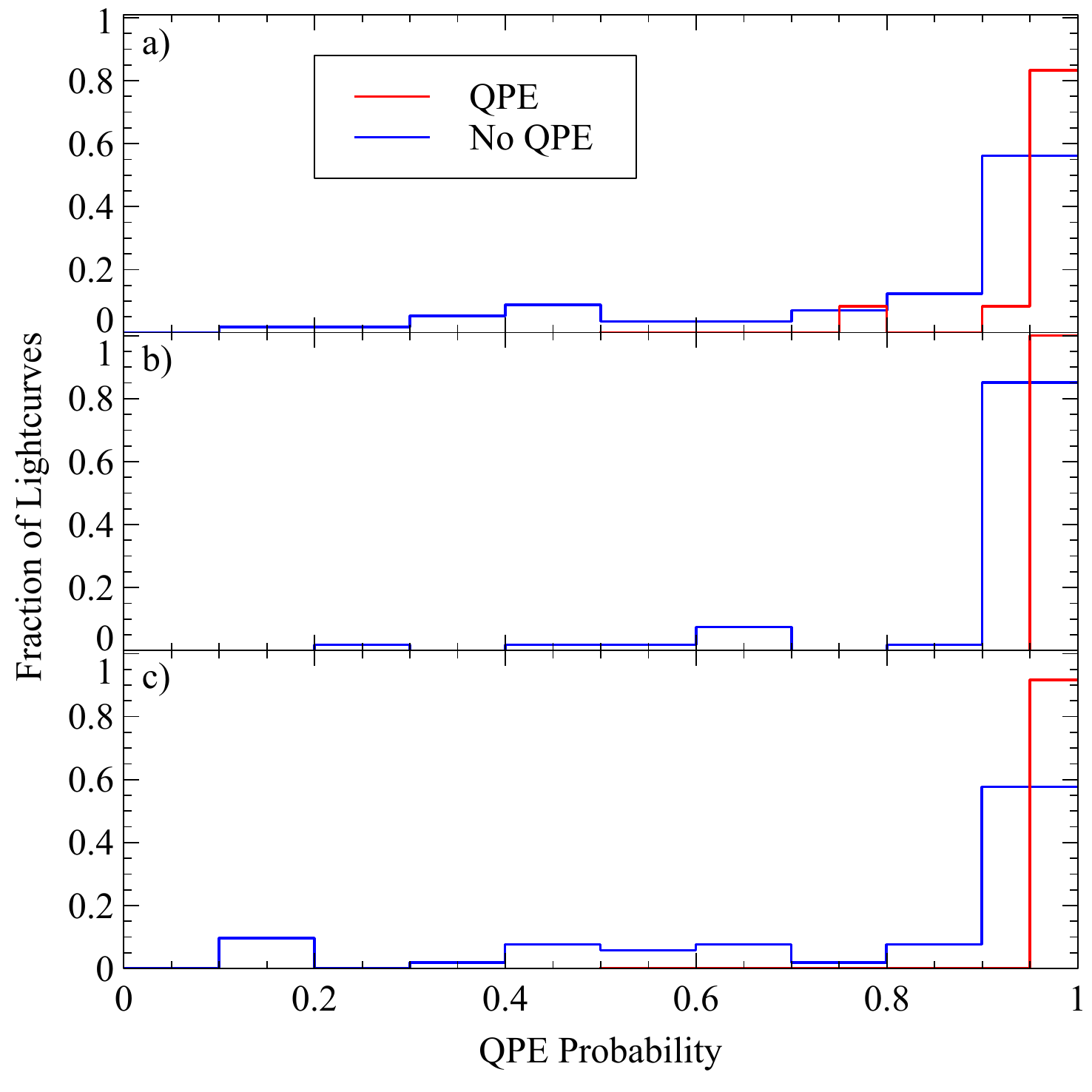}
    \caption{Probability that individual lightcurves in the observational data sample contain QPEs as per the results of the neural network classification. Panels show the results for lightcurves binned at a) 50s, b) 250s, and c) 1000s. The QPE and no QPE samples displayed are normalised by the number of lightcurves in each sample.}
    \label{fig:class-probs}
\end{figure}

\section{Discussion}
\label{sec:discuss}

\subsection{Simulated Lightcurve Feature Distributions}
\label{subsec:simlcfeatdists}

\begin{figure*}
	\includegraphics[width=0.9\textwidth]{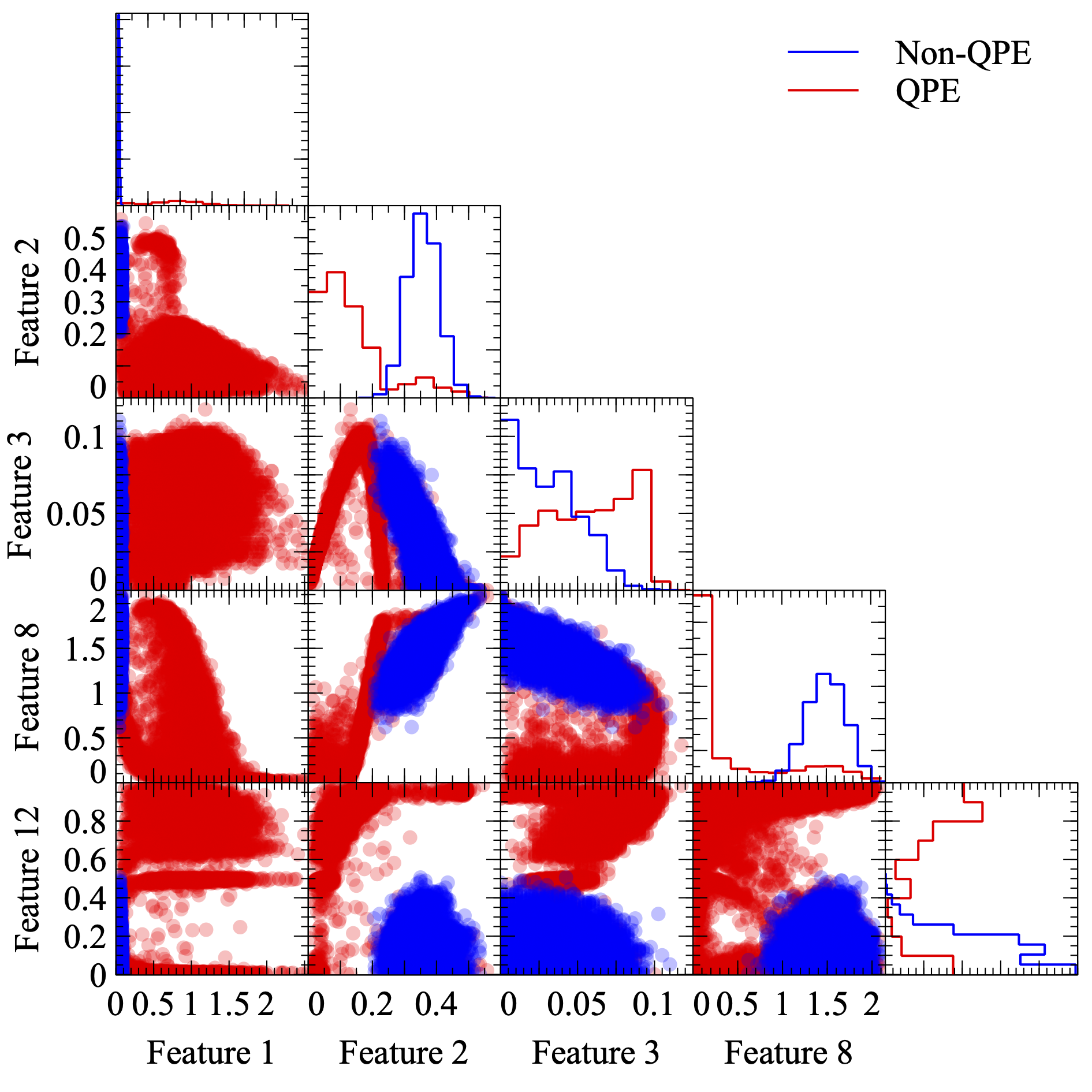}
    \caption{Distribution of feature values for features 1, 2, 3, 8 and 12. For clarity the features for a sample of 10000 of each of the simulated populations of lightcurves are shown in red (QPE) and blue (Non-QPE). Histograms show the frequency density distribution of features for all 50000 lightcurves in each class.}
    \label{fig:dt250-cornerplt} 
\end{figure*}

The simulated lightcurves were created according to a well-established algorithm which has been used for many purposes \citep[e.g.][etc.]{panagiotou_explaining_2022,hubner_pitfalls_2022}, and several of the features selected were chosen to match approaches taken in other attempts to identify variability in time-domain data \citep[e.g.][etc.]{sokolovsky_comparative_2017,pashchenko_machine_2018,faisst_how_2019}. None of the features were, however, selected with the expectation that they would serve this specific purpose in the identification of lightcurves containing quasi-periodic eruptions.
\begin{table}
	\centering
	\caption{Average and spread on feature values for the 14 features as described in section \ref{subsec:var_params}. Values stated are the mean and $\sigma$ from all 50,000 simulated lightcurves used in the training, validation and testing phases for each of the three values for time binning in the samples with and without imposed QPE signals.}
	\label{tab:feat-dists}
	\begin{tabular}{lccr} 
		\hline
		Feature & $\Delta T$ & QPE sample & Non-QPE sample \\
		\hline
            \multirow{3}{4em}{$f_{\text{1}}$} & 50 & $0.881\pm0.449$ & $0.047\pm0.009$ \\
             & 250 & $0.883\pm0.448$ & $0.048\pm0.010$ \\
             & 1000 & $0.878\pm0.449$ & $0.051\pm0.010$ \\
             \hline
            \multirow{3}{4em}{$f_{\text{2}}$} & 50 & $0.134\pm0.105$ & $0.358\pm0.047$ \\
             & 250 & $0.134\pm0.105$ & $0.358\pm0.048$ \\
             & 1000 & $0.135\pm0.106$ & $0.359\pm0.051$ \\
             \hline
            \multirow{3}{4em}{$f_{\text{3}}$} & 50 & $0.057\pm0.028$ & $0.029\pm0.021$ \\
             & 250 & $0.057\pm0.028$ & $0.029\pm0.021$ \\
             & 1000 & $0.057\pm0.030$ & $0.029\pm0.023$ \\
             \hline
            \multirow{3}{4em}{$f_{\text{4}}$} & 50 & $0.025\pm0.018$ & $0.0003\pm0.0016$ \\
             & 250 & $0.025\pm0.018$ & $0.0003\pm0.0016$ \\
             & 1000 & $0.025\pm0.019$ & $0.0002\pm0.0018$ \\
             \hline
            \multirow{3}{4em}{$f_{\text{5}}$} & 50 & $0.011\pm0.011$ & $(2.0\pm223.6)\times10^{-7}$ \\
             & 250 & $0.011\pm0.011$ & $(5.0\pm474.3)\times10^{-7}$ \\
             & 1000 & $0.011\pm0.012$ & $(4.0\pm632.4)\times10^{-7}$ \\
             \hline
            \multirow{3}{4em}{$f_{\text{6}}$} & 50 & $0.0049\pm0.0071$ & $0.0\pm0.0$ \\
             & 250 & $0.0049\pm0.0071$ & $0.0\pm0.0$ \\
             & 1000 & $0.0049\pm0.0078$ & $0.0\pm0.0$ \\
             \hline
            \multirow{3}{4em}{$f_{\text{7}}$} & 50 & $0.0025\pm0.0047$& $0.0\pm0.0$ \\
             & 250 & $0.0025\pm0.0047$ & $0.0\pm0.0$ \\
             & 1000 & $0.0026\pm0.0053$ & $0.0\pm0.0$ \\
             \hline
            \multirow{3}{4em}{$f_{\text{8}}$} & 50 & $0.445\pm0.574$ & $1.50\pm0.22$ \\  
             & 250 & $0.445\pm0.572$ & $1.49\pm0.22$ \\
             & 1000 & $0.448\pm0.570$ & $1.48\pm0.23$ \\
             \hline
            \multirow{3}{4em}{$f_{\text{9}}$} & 50 & $3.24\pm2.38$ & $-0.0012\pm0.3978$ \\
             & 250 & $3.24\pm2.37$ & $0.002\pm0.399$ \\
             & 1000 & $3.16\pm2.24$ & $-0.0029\pm0.4044$ \\
             \hline
            \multirow{3}{4em}{$f_{\text{10}}$} & 50 & $15.8\pm25.0$ & $-0.63\pm0.49$ \\
             & 250 & $15.7\pm24.4$ & $-0.626\pm0.491$ \\
             & 1000 & $14.4\pm19.7$ & $-0.637\pm0.504$ \\
             \hline
            \multirow{3}{4em}{$f_{\text{11}}$} & 50 & $1.33\pm700.05$ & $5.9\pm913.9$ \\
             & 250 & $1.12\pm141.10$ & $0.066\pm182.072$ \\
             & 1000 & $-0.44\pm34.89$ & $0.22\pm46.05$ \\
             \hline
            \multirow{3}{4em}{$f_{\text{12}}$} & 50 & $0.62\pm0.34$ & $0.12\pm0.09$ \\
             & 250 & $0.616\pm0.337$ & $0.124\pm0.088$ \\
             & 1000 & $0.575\pm0.318$ & $0.125\pm0.089$ \\
             \hline
            \multirow{3}{4em}{$f_{\text{13}}$} & 50 & $0.98\pm0.04$ & $0.97\pm0.02$ \\
             & 250 & $0.914\pm0.096$ & $0.922\pm0.039$ \\
             & 1000 & $0.731\pm0.216$ & $0.839\pm0.077$ \\
             \hline
            \multirow{3}{4em}{$f_{\text{14}}$} & 50 & $0.017\pm0.125$ & $0.0049\pm0.0031$ \\
             & 250 & $0.176\pm0.308$ & $0.027\pm0.017$ \\
             & 1000 & $1.15\pm0.68$ & $0.114\pm0.068$ \\
		\hline
	\end{tabular}
\end{table}
When we look at the features being extracted from the simulated lightcurves we can see that there are several at all time binning rates which appear to distinguish well between lightcurves with and without QPEs. Features 1, 2, 4, 8, 9, and 12 across all time binning rates are separated by at least 1$\sigma$, as shown in Table \ref{tab:feat-dists}. It should be noted that all of the features except 9 and 11 are constrained to be positive only. In some of the cases the feature distributions cannot be used in all cases to classify the lightcurve as containing a QPE or not, but specific values may be strong indicators as containing a QPE or not. In the cases of features 6 and 7, any non-zero value is strongly indicative of a lightcurve containing QPEs, but not all QPE containing lightcurves have non-zero values for these features. The values for feature 1, which appear to be a good way of distinguishing between the two classes, should also be taken with a measure of caution. They are affected by the factor which shifts the lightcurves from being zero-centred (a necessary step for the imposition of the QPE signals in the lightcurve generation), and a further investigation would be needed in order to determine what effect varying this factor has on the distributions for feature 1 in the two classes and whether it remains a distinguishing factor.
\begin{figure*}
	\includegraphics[width=0.98\textwidth]{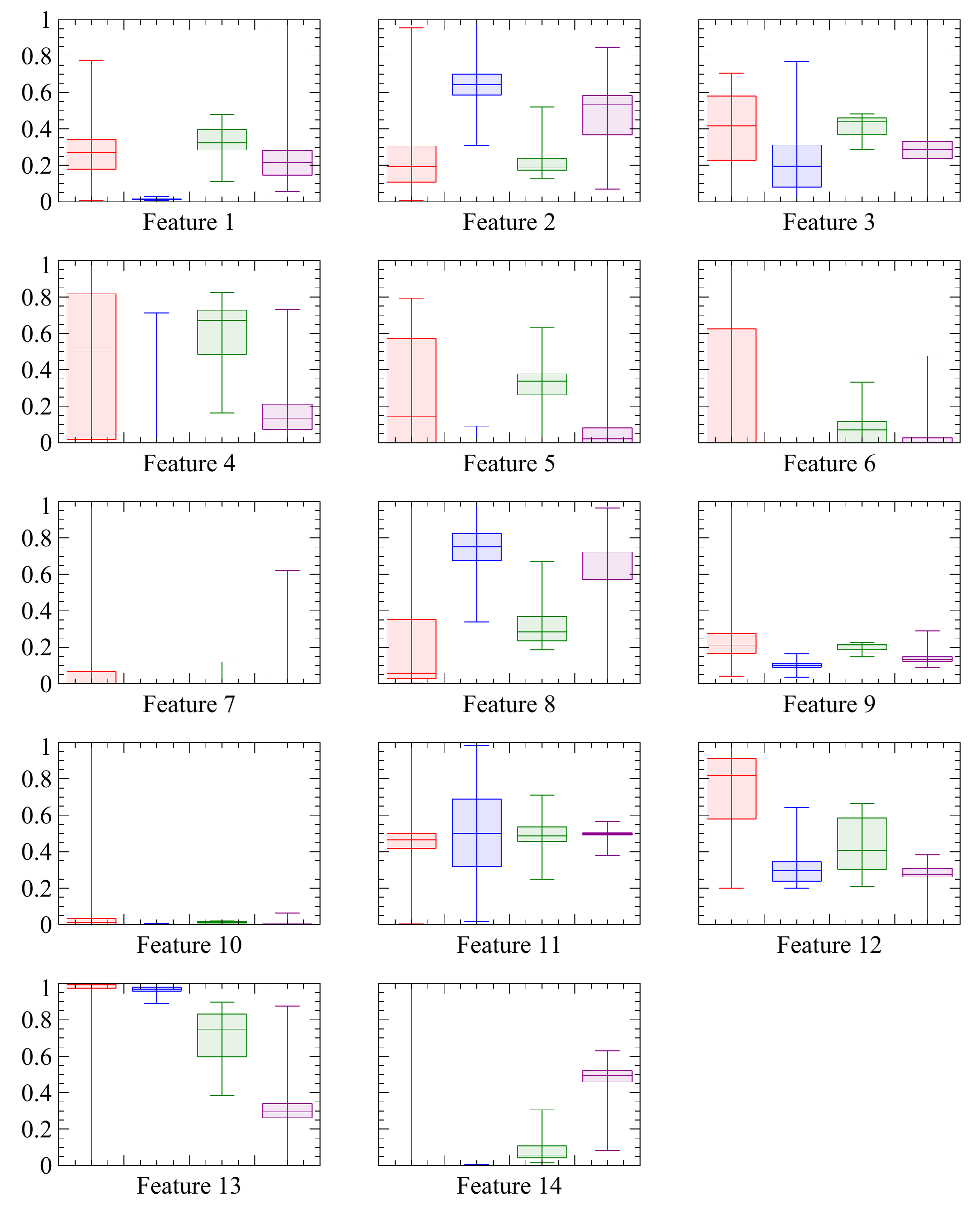}
    \caption{Distribution of feature values for the 14 features used in the classification derived from lightcurves binned at 50s. Box plots display the distribution of features from left to right for the populations of: simulated lightcurves containing QPEs (red); simulated lightcurves without QPEs (blue); observed lightcurves containing QPEs (green); observed lightcurves without QPEs (purple). Feature values have been normalized between 0 and 1.}
    \label{fig:dt50-featdists}
\end{figure*}
\begin{figure*}
	\includegraphics[width=0.98\textwidth]{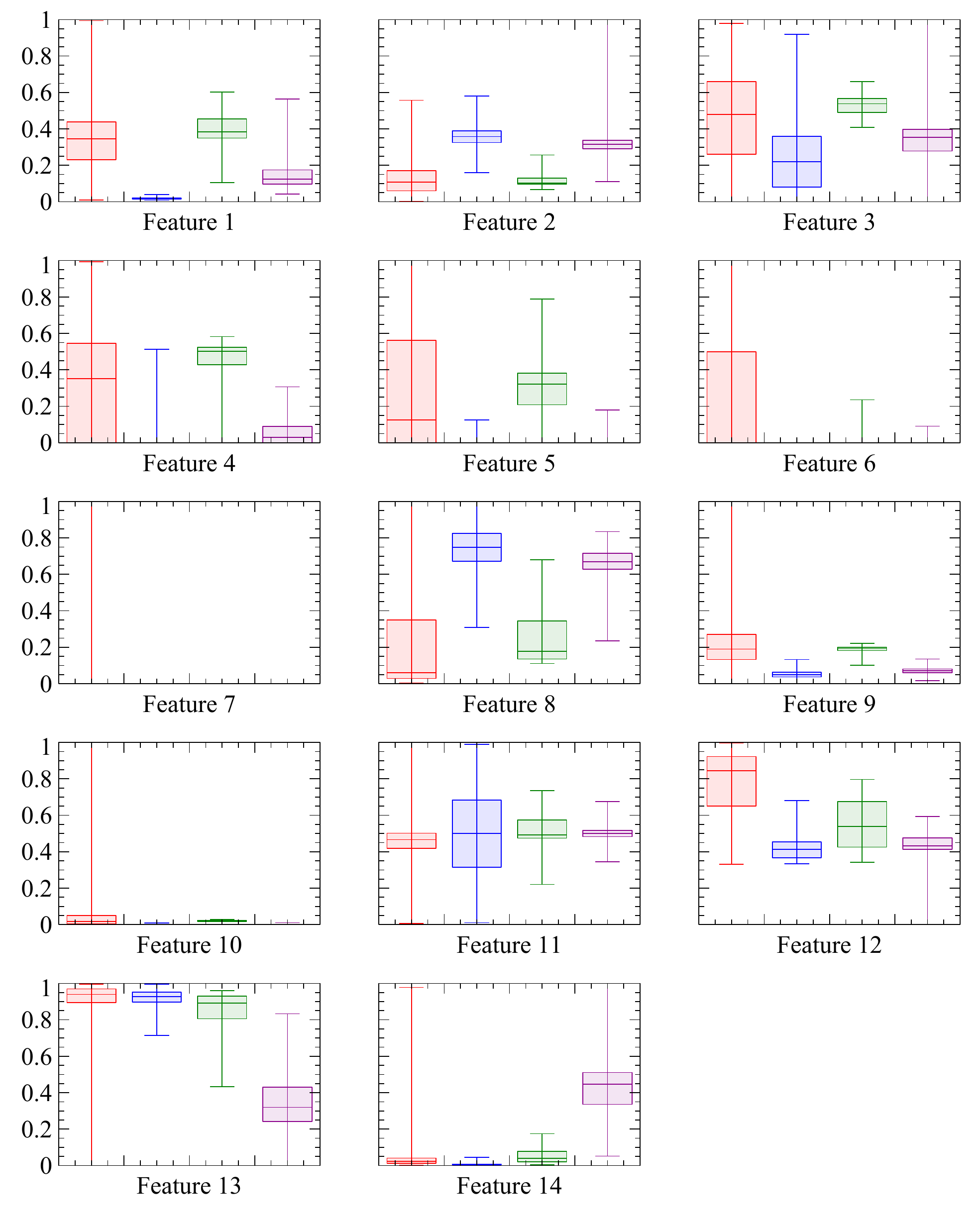}
    \caption{Distribution of feature values for the 14 features used in the classification derived from lightcurves binned at 250s. Box plots display the distribution of features from left to right for the populations of: simulated lightcurves containing QPEs (red); simulated lightcurves without QPEs (blue); observed lightcurves containing QPEs (green); observed lightcurves without QPEs (purple). Feature values have been normalized between 0 and 1.}
    \label{fig:dt250-featdists}
\end{figure*}
\begin{figure*}
	\includegraphics[width=0.98\textwidth]{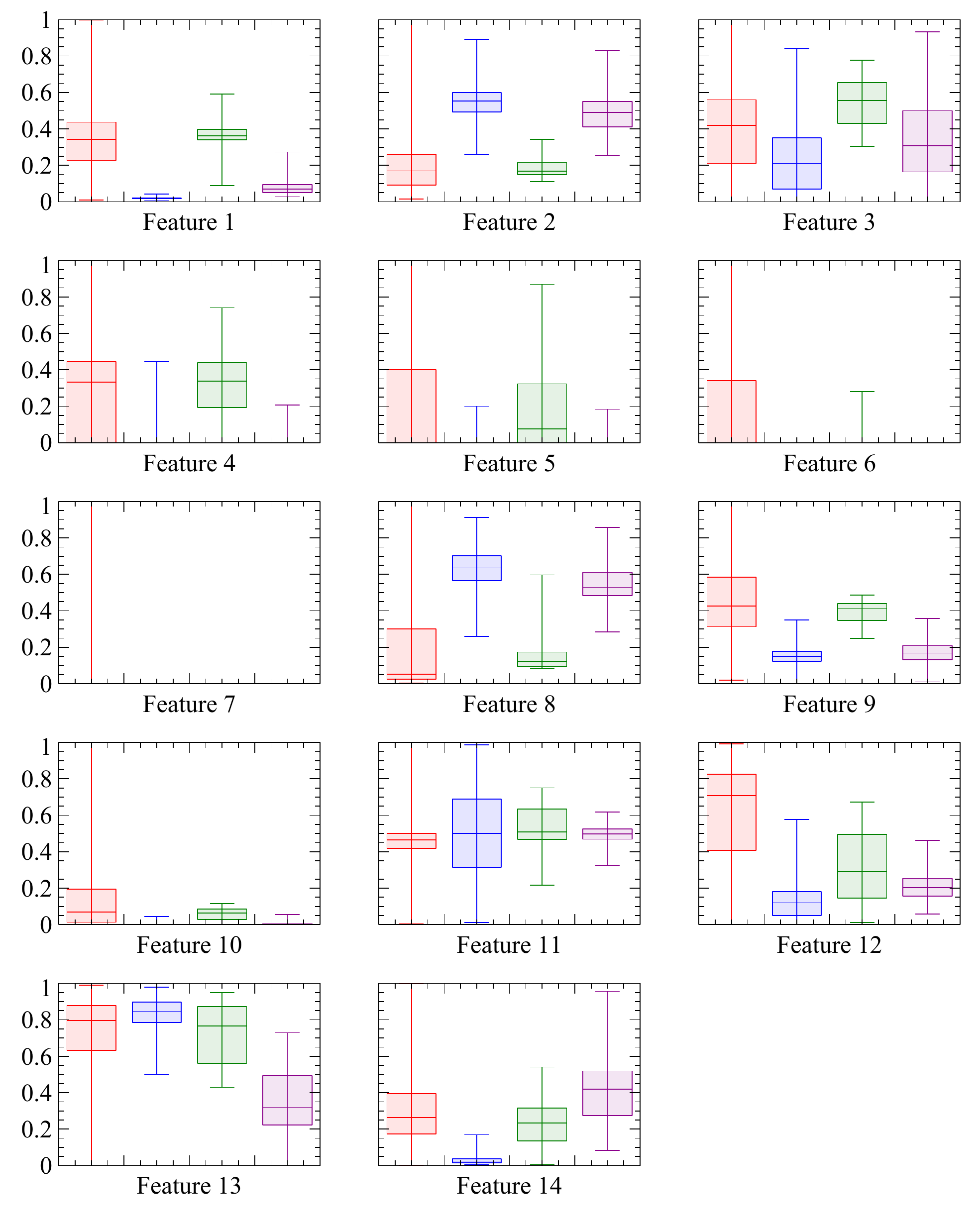}
    \caption{Distribution of feature values for the 14 features used in the classification derived from lightcurves binned at 1000s. Box plots display the distribution of features from left to right for the populations of: simulated lightcurves containing QPEs (red); simulated lightcurves without QPEs (blue); observed lightcurves containing QPEs (green); observed lightcurves without QPEs (purple). Feature values have been normalized between 0 and 1.}
    \label{fig:dt1000-featdists}
\end{figure*}
If we consider the distributions of the features, as shown in Fig. \ref{fig:dt50-featdists}, \ref{fig:dt250-featdists}, and \ref{fig:dt1000-featdists} we can see that for several of the features the populations show differences, both for the simulated and observed samples. Features 1, 2, 3, 8, and 12 all show distinct differences between the QPE and non-QPE populations, with the distributions for the real observed lightcurves and the simulated lightcurves in each class clearly overlapping for all three time bin values. Fig. \ref{fig:dt250-cornerplt} shows how these five features interact in more detail.
From this we can see that the combinations of $f_\text{1}$ and $f_\text{2}$, $f_\text{2}$ and $f_\text{3}$, $f_\text{2}$ and $f_\text{12}$, and $f_\text{8}$ and $f_\text{12}$ appear to show a high level of separation between the two classes of simulated lightcurves.

Kolmogorov-Smirnov testing for all fourteen features reveals that for the simulated feature distributions the classes are distinct at the $1\%$ confidence level for each individual feature at all time bin values. In comparing the feature distributions for the two classes of real observations features 1, 2, 4, 8, 9, 10 and 13 were individually significantly different at the $1\%$ confidence level for all time bin values. Features 3, 5 and 14 were all individually significantly different at a confidence level of $5\%$ for all three time bin values, and feature 12 was significant at $10\%$. Features 6 and 11 were also individually significantly different at the $10\%$ level for time binning of 50s, but were not significantly different at the other two time bin values. Feature 7 did not show a significant difference at any value for time binning. Features 6 and 7 are, however, affected by the time binning of the simulated lightcurves, as all simulated lightcurves are of duration 100ks, and so the number of points in the lightcurves decreases by a factor of twenty as the time binning increases from 50s to 1000s. Additionally, these results with regards to the observational feature distributions should be taken with some caution due to the reduced sample size; the QPE class only contains 12 lightcurves binned at 1000s.

If we consider networks trained for only individual features, and with architecture optimised as described in section \ref{subsec:nn-arch}, we find that certain features at different values for the time bin appear to be more informative than others. With a time binning of 50\,s we achieve a classification accuracy of 0.870 and $F_{\text{1}}$ score of 0.609 using only feature 3. With a time binning of 250\,s we achieve a classification accuracy of 0.970 and $F_{\text{1}}$ score of 0.917 using only feature 8, or an accuracy of 0.894 and $F_{\text{1}}$ score of 0.588 using only feature 12. With a time binning of 1\,ks we achieve a classification accuracy of 0.922 and $F_{\text{1}}$ score of 0.828 using only feature 9.

\subsection{Threshold Probability Optimisation}
\label{subsec:confopt}
Our default measure for distinguishing between lightcurves containing QPEs and not containing QPEs was to compare the classification probabilities, as the output of a \texttt{softmax} layer, and assign the class as that which has the highest probabliity. With our problem being that of a binary classification this means that initially we set a nominal threshold probability of 50$\%$ between the two classes. Following the results of the classification of the observational data in section \ref{subsec:reallcres} we then considered whether an adjustment to the level at which we split the two classes could improve the performance. We perform a simple grid search for threshold probability between 0 and 1 at steps of 0.000001 with the results of the classification at all three neural networks and find that altering the level at which we delineate between lightcurves containing QPEs and those which do not can significantly improve the performance of the classifier. The full results of the optimisation are given in Table \ref{tab:conf-opt}.
\begin{table}
	\centering
	\caption{Optimal ranges for the cut in probability value at which we define a lightcurve as containing QPEs or not. Threshold probability ranges are closed intervals which give the optimised threshold probability values for accuracy, purity, completeness and $F_{\text{1}}$ score.}
	\label{tab:conf-opt}
	\begin{tabular}{cccr} 
		\hline
		Time Bin (s) & Metric & Conf. Range & Metric Value \\
		\hline
            \multirow{4}{4em}{50} & Accuracy & [0.999206, 0.999211] & 0.8841 \\
             & Purity & [0.999206, 0.999211] & 0.6429\\
             & Completeness & [0, 0.778044] & 1.0 \\
             & $F_{\text{1}}$ Score & [0.999206, 0.999211] & 0.6923\\
             \hline
             \multirow{4}{4em}{250} & Accuracy & [0.999886, 0.999944] & 0.8939 \\
             & Purity & [0.999977, 0.999998] & 1.0 \\
             & Completeness & [0.0, 0.986412] & 1.0 \\
             & $F_{\text{1}}$ Score & [0.999886, 0.999893] & 0.6666\\
             \hline
             \multirow{4}{4em}{1000} & Accuracy & [0.999330, 0.999455] & 0.9844 \\
             & Purity & [0.999330, 1.0] & 1.0\\
             & Completeness & [0.0, 0.999999] & 1.0 \\
             & $F_{\text{1}}$ Score & [0.999330, 0.999455] & 0.9565\\
		\hline
	\end{tabular}
\end{table}
By varying the threshold probability at which we distinguish between QPE and non-QPE containing lightcurves we were able to make significant improvements in the accuracy and $F_{\text{1}}$ scores for all three data sets.
\begin{figure}
	\includegraphics[width=\columnwidth]{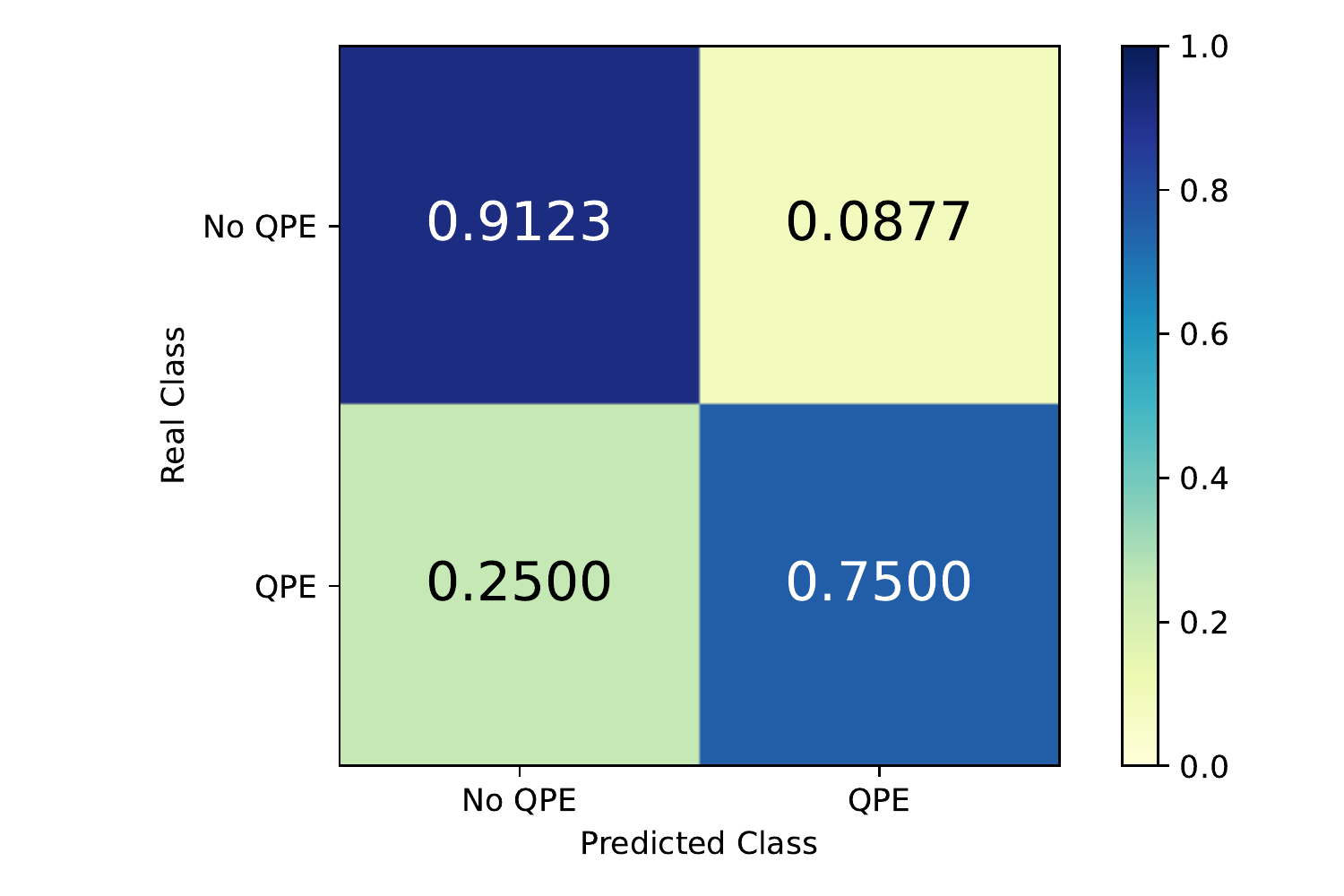}
    \caption{Confusion matrix showing the classification of 57 lightcurves without QPEs and 12 with lightcurves with QPEs present. The lightcurves were binned at a rate of 50s, and their full details are given in \ref{tab:all-obs1}. Threshold probability for requiring a QPE is set at the level which optimises accuracy of 0.99921.}
    \label{fig:dt50-optconfmat}
\end{figure}
In the case of all three classifiers we can improve the accuracy by raising the required QPE probability. For each of the three classifiers the optimal QPE probability was very close to 1.0, which is indicated by the distributions of lightcurve probabilities as seen in Fig. \ref{fig:class-probs}.
\begin{figure}
	\includegraphics[width=\columnwidth]{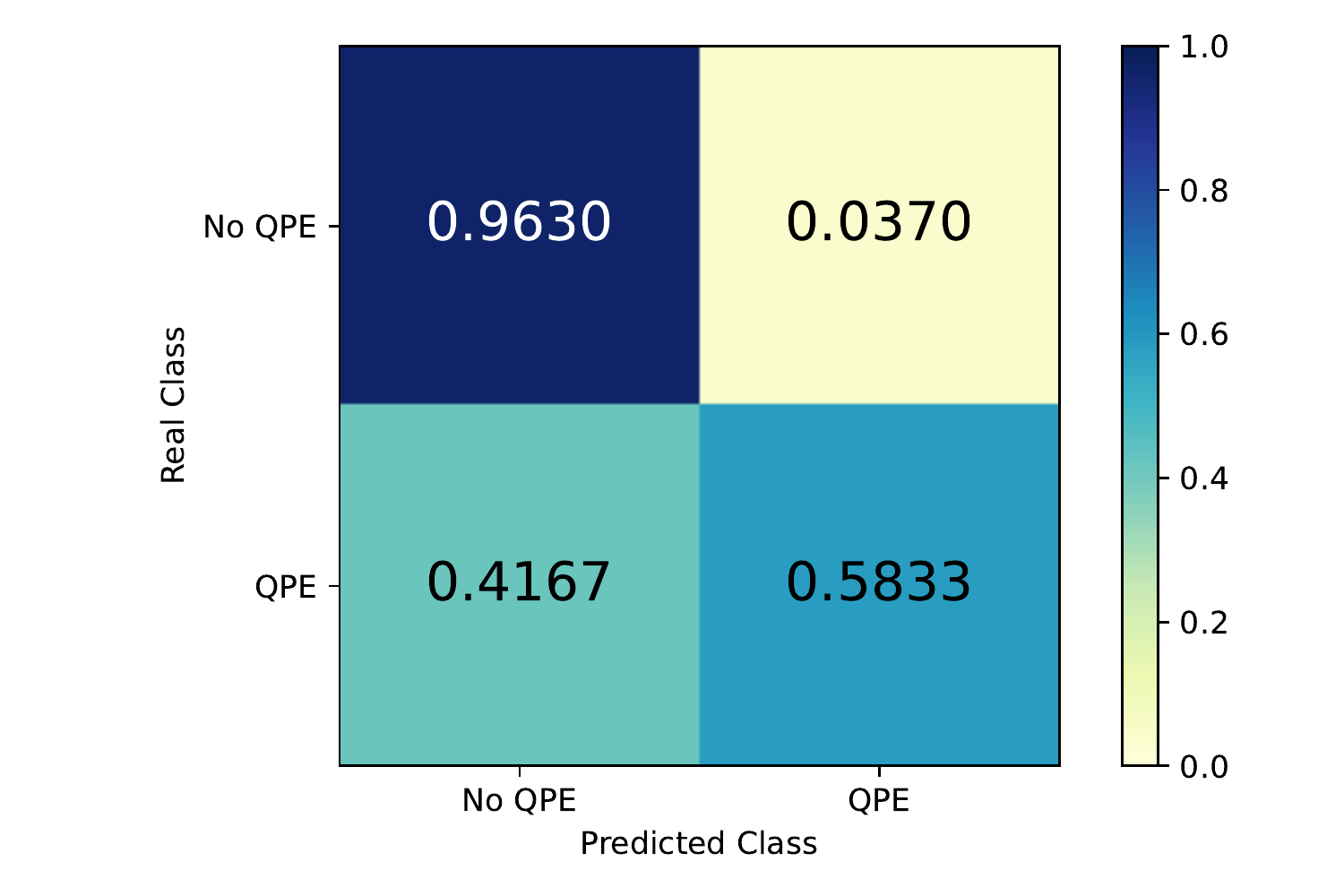}
    \caption{Confusion matrix showing the classification of 54 lightcurves without QPEs and 12 with lightcurves with QPEs present. The lightcurves were binned at a rate of 250s, and their full details are given in \ref{tab:all-obs1}. Threshold probability for requiring a QPE is set at the level which optimises accuracy of 0.9999.}
    \label{fig:dt250-optconfmat}
\end{figure}
In all three cases by optimising the accuracy we reduce the completeness of the classifier, but given that the sample being classified is biased towards lightcurves not containing QPEs the increase in false negative classifications is far outweighed by the decreases in each case of false positive classifications. For the 50s classifier 4 out of the 12 QPE containing lightcurves are misclassified at a higher probability, but an additional 46 of the 57 non-QPE lightcurves are then correctly classified, and we see a similar increase in the performance of the 250s classifier.
\begin{figure}
	\includegraphics[width=\columnwidth]{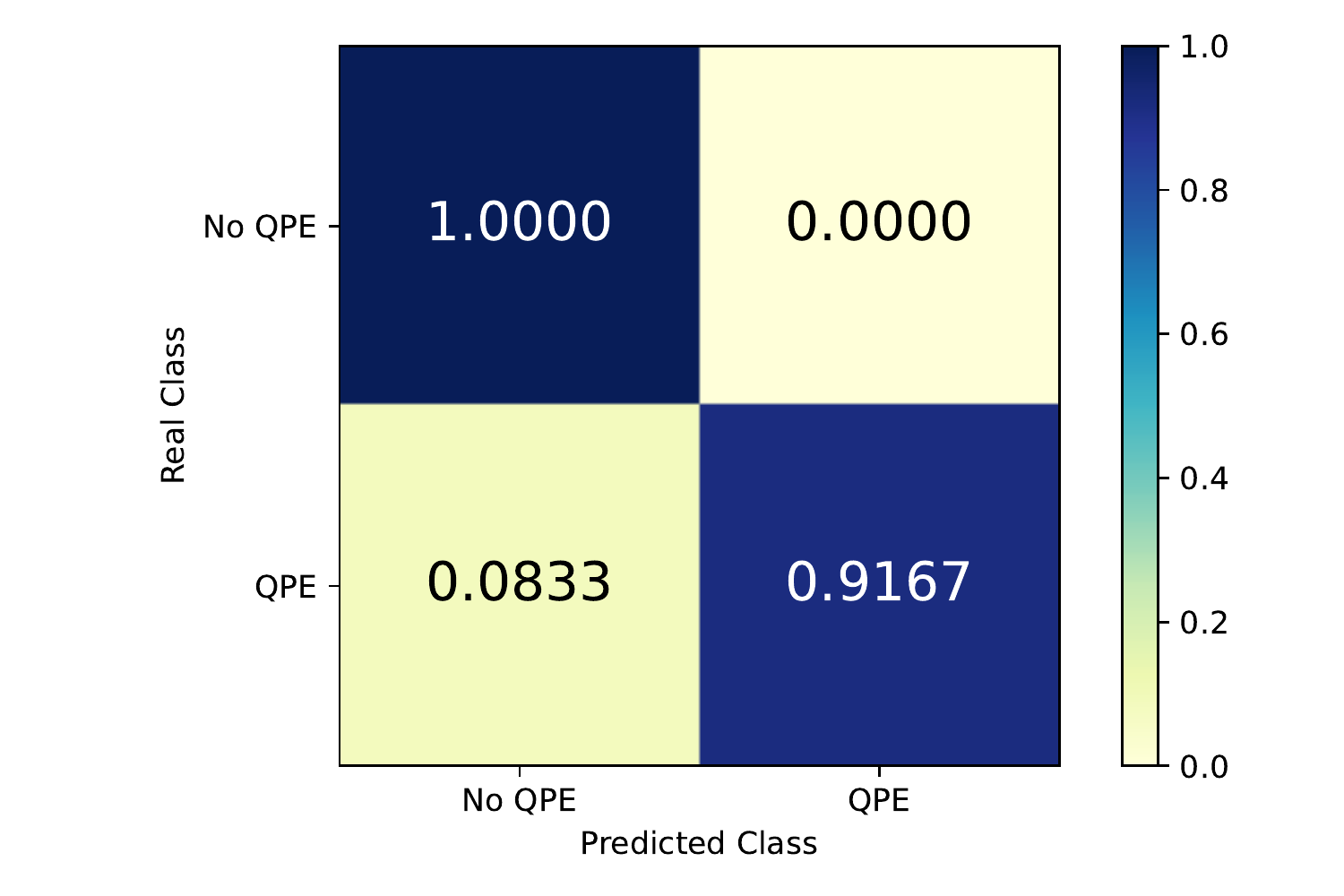}
    \caption{Confusion matrix showing the classification of 52 lightcurves without QPEs and 12 with lightcurves with QPEs present. The lightcurves were binned at a rate of 1000s, and their full details are given in \ref{tab:all-obs1}. Threshold probability for requiring a QPE is set at the level which optimises accuracy of 0.9994.}
    \label{fig:dt1000-optconfmat}
\end{figure}
The greatest increase in performance is seen in the 1000s classifier, where at the optimal threshold probability level for accuracy we reduce the mis-classification rate for non-QPE containing lightcurves to $0\%$.

\begin{figure}
	\includegraphics[width=\columnwidth]{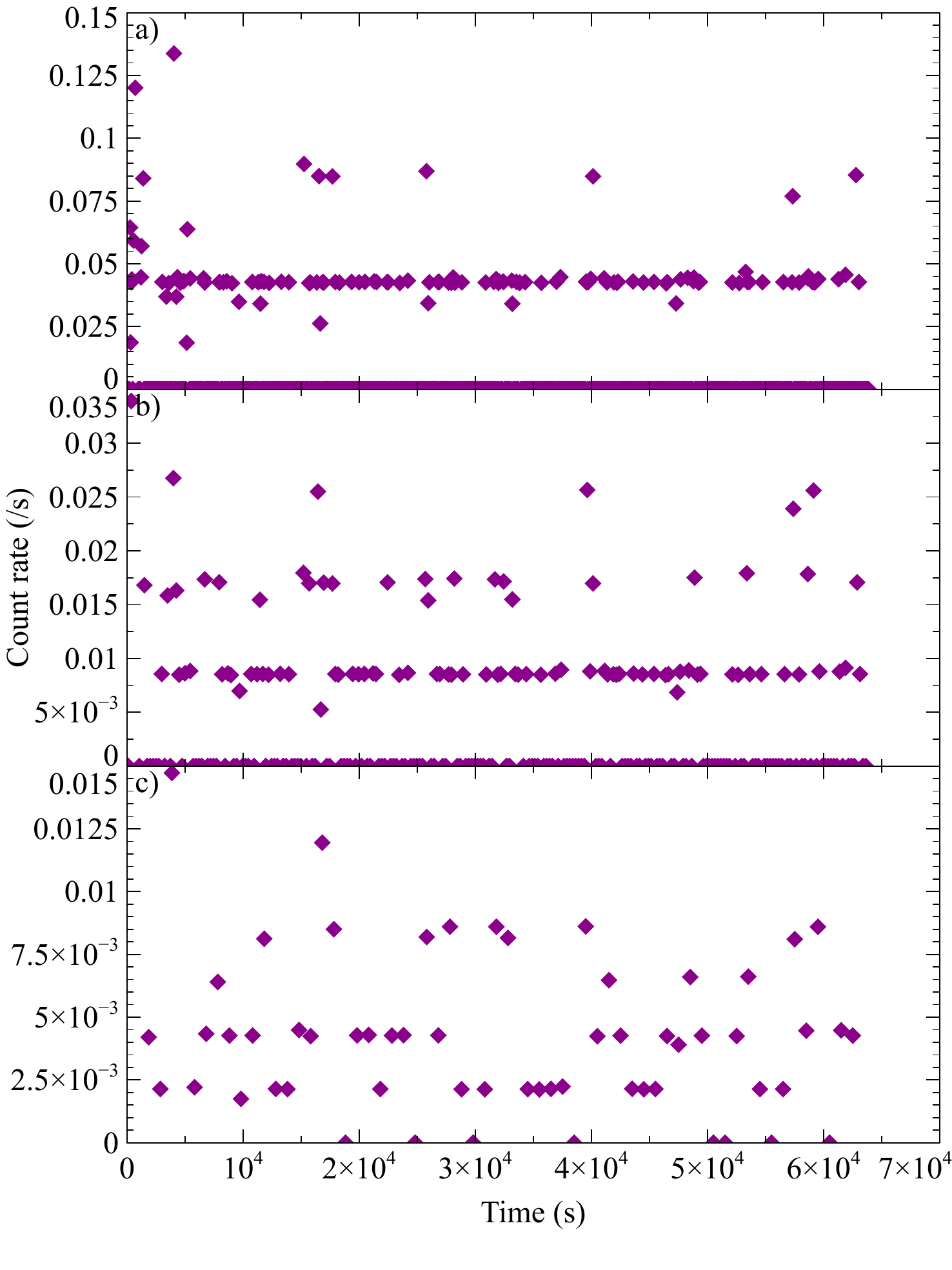}
    \caption{Lightcurve for observation 0304190101 of NGC 1331 which was misclassified as containing QPEs when binned at 50s and 250s. The lightcurve is shown binned at rates of a) 50s, b) 250s, and c) 1000s and was misclassified in all cases. It appears likely that the misclassification is due to the effects of Poisson noise being more amplified given the low average count rate, which would be most noticeable for the two shorter time bin values.}
    \label{fig:class-falsepos}
\end{figure}

\begin{figure}
	\includegraphics[width=\columnwidth]{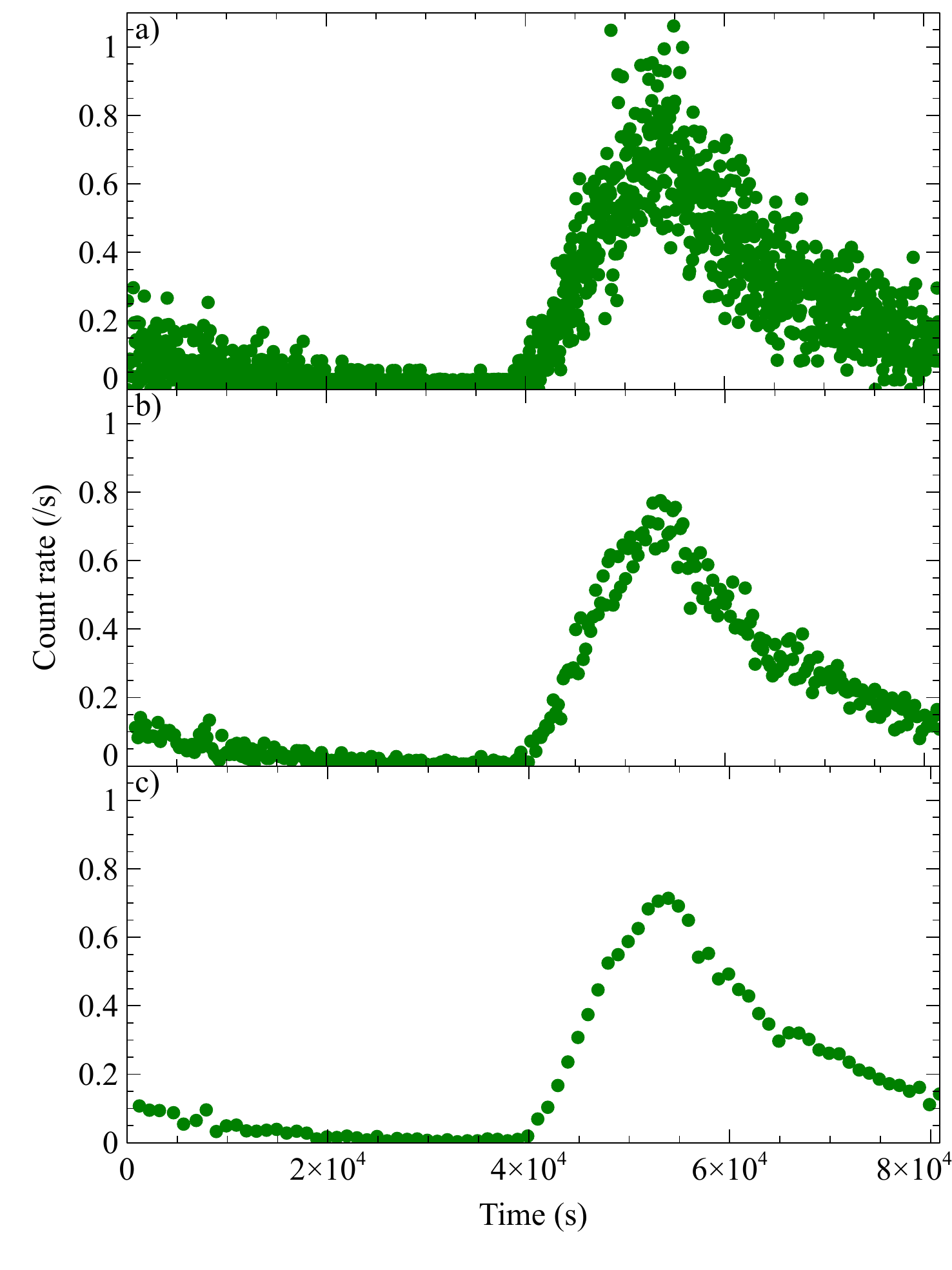}
    \caption{Lightcurve for observation 0861910301 of eRASSU J023147.2-102010 which was misclassified as not containing QPEs. The lightcurve is shown binned at rates of a) 50s, b) 250s, and c) 1000s and was misclassified in all cases.}
    \label{fig:class-falseneg}
\end{figure}

After optimisation with all three classifiers there is one false negative result which appears in all three samples, and one false positive which appears in both the 50s and 250s binned data sets. Observation 0304190101 of NGC 1331 is classed as containing a QPE with probability 0.999999, 0.999976, and 0.999329 across the 50s, 250s and 1000s classifiers respectively. The lightcurve for observation 0304190101 of NGC 1331 is shown in Fig. \ref{fig:class-falsepos}. Observation 0861910301 of eRASSU J023147.2-102010 is classed as not containing a QPE with probability 0.986134, 0.995476, and 0.318089 across the 50s, 250s and 1000s classifiers respectively. The lightcurve for observation 0861910301 of eRASSU J023147.2-102010 is shown in Fig. \ref{fig:class-falseneg}.


If we look at the individual features, extracted from the 250s binned lightcurves, which achieved the highest levels of classification accuracy, we find that in both cases many of the features for the erroneously classified lightcurves are more indicative of their incorrect classes. For observation 0304190101 of NGC 1331 features 1, 9 and 10 were outliers for the Non-QPE class, with features 1 and 9 showing significant differences between the two classes. For values of $f_\text{1}=1.52$ and $f_\text{9}=1.59$ this observation sits at 1.42$\sigma$ (147.2$\sigma$) and 0.70$\sigma$ (3.98$\sigma$) from the mean for the QPE (Non-QPE) class. The average deviation from the class feature means for observation 0304190101 was 1.50$\sigma$ for the QPE class, and 34.51$\sigma$ for the Non-QPE class. For observation 0861910301 of eRASSU J023147.2-102010 features 4, 5, 6, 7, 8, 12 and 14 were outliers for the QPE class, with features 4, 5, 6, 8 and 12 being the most extreme values for the QPE class. For values of $f_\text{4}=0.0$, $f_\text{5}=0.0$, $f_\text{8}=1.36$, and $f_\text{12}=1.36$ this observation sits at $0.156\sigma$ ($1.34\sigma$), $0.011\sigma$ ($0.935\sigma$), $0.060\sigma$ ($1.605\sigma$), and $1.26\sigma$ ($1.79\sigma$) from the mean for the Non-QPE (QPE) class. The average deviation from the class feature means for observation 0861910301 was 0.858$\sigma$ for the QPE class, and 10.22$\sigma$ for the Non-QPE class. As the features overall for both of these observations would seem to indicate that both share more in common with the QPE class it is evident that the effects of individual features which may be more important than others are causing the misclassifications.

A particular issue in this classification is the high rate of false positive identifications, as QPEs are rare in appearance. If we look at those lightcurves which do not contain QPEs we see a distinct difference in the average count rates for those which are correctly or incorrectly identified across all three time bin values. Those lightcurves which do not contain QPEs have average count rates of 0.402, 0.439 and 0.447 $\text{\,cts\,s}^{-1}$ across the three time bin values when considering those which are available for classification. The average count rates for those observations which are categorised correctly (and incorrectly) are 0.472 (0.051) for the 50s sample, 0.447 (0.005) for the 250s sample, and 1.045 (0.103) $\text{\,cts\,s}^{-1}$ for the 1000s sample. Observation 0304190101, shown in Fig. \ref{fig:class-falsepos}, is misclassified with all three networks and has an average count rate of $0.0047 \text{\,cts\,s}^{-1}$, indicating that even when binned at 250s it is likely that there will be a high proportion of empty bins and as such the classification may well be affected by Poisson noise due to the low count rates.

\subsection{Application to the XMM Serendipitous Source Catalogue}
\label{subsec:xmmssc_res}
Having achieved high levels of accuracy in classifying simulated lightcurves, and with the moderate level of success achieved in classifying real observational data, we now consider whether any new QPE candidates can be identified through application of the networks to the XMM Serendipitous Source Catalogue. 

We filter the XMM Serendipitous Source Catalogue (4XMM-DR12)\footnote{\url{http://xmmssc.irap.omp.eu/Catalogue/4XMM-DR12/4XMM_DR12.html}} for observations which contain pn detections, have lightcurve data which are created as described in \cite{webb_xmm-newton_2020}, and have total exposure times greater than 50ks. Detections are referred to by their 'SRCID' tags as given in the 4XMM-DR12 data release. We also screen for any detections which could be spurious (possibly near an extended bright source, or where the detector coverage is particularly low) using the criteria that {\tt SUM\_FLAG=0}. From the total catalogue of 939,270 items this creates a sample of 83,531 detections for consideration. We download the pn lightcurve data for these detections, and then screen these lightcurves for time bins of less than 50s, and rebin the lightcurve to 50s, 250s and 1000s. Due to issues with very low count rate observations as identified in section \ref{subsec:confopt} we then screen for any lightcurves with an average count rate below $0.01 \text{\,cts\,s}^{-1}$. Finally, due to issues with flaring events which are common at the start or end of observations depending on their position during observing cycles we then remove the first and last 15ks of each observation. Features are then extracted from each of the three lightcurves for each detection and the objects are classified according to the networks as trained in section \ref{subsec:nn-arch}, and tested on the observational data as described in section \ref{subsec:reallcres}. 
As the phenomena we are trying to find is very rare we want to reduce the numbers of false positive events in any sample which is to be scheduled for follow-up by an expert observer. In Fig. \ref{fig:dt50-optconfmat}, \ref{fig:dt250-optconfmat} and \ref{fig:dt1000-optconfmat} we can see that with very high threshold probabilities we could still flag very large numbers of false positive results given the total sample size. As such, we select detections from the catalogue for further review based on the result of classification from all three networks. In order for a lightcurve to be flagged for further review we require that it achieve a score of:
\begin{itemize}
\item 0.999 with features extracted from a 50s binned lightcurve.
\item 0.9999 with  features extracted from a 250s binned lightcurve.
\item 0.999 with features extracted from a 1000s binned lightcurve.
\end{itemize}
At these threshold probabilities we achieve purity (and completeness) scores of 0.6 (0.75), 0.86 (0.58), and 0.85 (0.92) with the 50s, 250s and 1000s classifiers respectively on our initial testing data, with $22\%$, $11\%$ and $20\%$ of the lightcurves in our sample being flagged at these threshold probability levels.
With these selection criteria from the total sample of 83,531 detections 705 lightcurves are identified for further examination with six being classified as containing QPEs with a probability of 1.0 across all three networks.

\begin{table*}
	\centering
	\caption{Predictions for the eleven observations of known QPE sources contained within the XMM Serendipitous Source Catalogue (4XMM-DR12). For each object we note the observation ID, the predictions from the three networks for the three time bin values (P50, P250 and P1000), and whether the lightcurve is flagged for further inspection as per the methodology in section \ref{subsec:xmmssc_res}.}
	\label{tab:qpeml-qpepreds}
	\begin{tabular}{lcccccr}
		\hline
		Object & OBSID & Eruptions & P50 & P250 & P1000 & Flag\\
		\hline
            GSN 069 & 0823680101 & 2 & 0.999 & 0.9999 & 1.000 & Y \\
            -- & 0831790701 & 5 & 0.9986 & 0.9993 & red{0.999} & N \\
            -- & 0851180401 & 5 & 0.999 & 0.9999 & 0.999 & Y \\
            -- & 0864330101 & 4 & 0.829 & 0.9772 & 0.996 & N \\
		RX J1301.9+2747 & 0124710801 & 1 & 0.99898 & 0.9995 & 1.000 & N \\
            -- & 0851180501 & 3 & 1.000 & 0.9995 & 1.000 & Y \\
            -- & 0864560101 & 8 & 0.998 & 0.9999 & 0.999 & N \\
            XMMSL1 J024916.6-041244 & 0411980401 & 1 & 0.768 & 0.9721 & 0.999 & N \\
            eRASSU J023147.2-102010 & 0861910201 & 2 & 0.99898 & 0.9998 & 0.984 & N \\
            -- & 0861910301 & 1 & 0.952 & 0.9767 & 0.161 & N \\
            eRASSU J023448.9-441931 & 0872390101 & 9 & 1.000 & 0.9999 & 1.000 & Y \\
		\hline
	\end{tabular}
\end{table*}
When this selection criteria is applied to the known observations of QPE candidates which are in the 4XMM-DR12 catalogue we find that for the 11 observations which are in the catalogue only four meet our selection criteria. Of those 7 observations which do not meet the criteria one only fails the criteria for one of the three networks. In all but one of the cases the predictions for the observations are strongly favoured towards containing QPEs, see Table \ref{tab:qpeml-qpepreds} for the full results. The very low score for observation 0861910301 of eRASSU J023147.2-102010 when binned at 1\,ks is an outlier. These results are not entirely unexpected due to the nature of the pre-processed lightcurves available with the XMM Serendipitous Source Catalogue. The lightcurves which are associated with the catalogue are for the full range of energies detected by \emph{XMM-Newton} (0.2--12.0keV), and we know from previous examination of QPE sources that the phenomena is largely contained to lower energies \citep[][etc.]{Miniutti2019,Giustini2020,Arcodia2021,Chakraborty2021}. As such, the variability which is clear and evident in the 0.2--2.0keV band appears suppressed and of a smaller amplitude when observed in the full 0.2--12.0\,keV range, and it is not surprising that fewer of the known QPE sources meet our criteria for flagging in this case. Across the different time bin values we observe changes in the features when considering the lightcurve for the full 0.2--12.0\,keV energy band. Feature 1 for all observations decreased with a change to the energy band, feature 9 displayed an average decrease across all three time bin values, and for 50s and 250s binned curves feature 8 displayed an average increase. All three of these changes are indicative of movement from QPE to non-QPE populations, and would explain the decreased probabilities of the classifications towards containing QPEs.

\begin{figure}
	\includegraphics[width=\columnwidth]{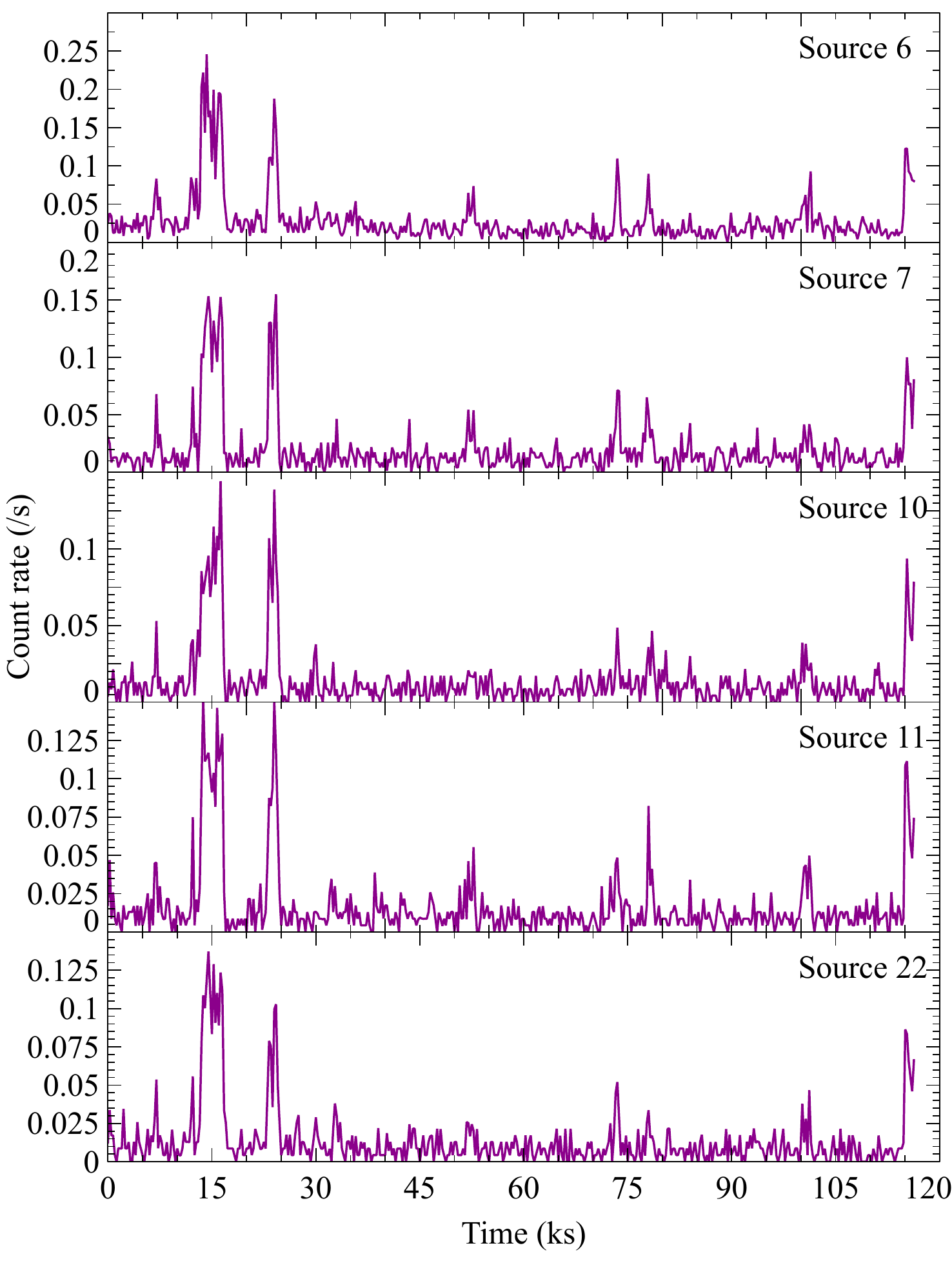}
    \caption{Lightcurves for five sources flagged as containing QPEs within the observation 0804680101. All lightcurves are binned at 250s and show a consistent pattern of variability despite sources being located on different sections of the detector. Panels display (from top to bottom) the lightcurve for sources 6, 7, 10, 11, and 22.}
    \label{fig:multi-det-obs-pattern}
\end{figure}

We then manually examined the lightcurves for the 705 detections which were flagged from the Serendipitous Source Catalogue. The first result of note is that a large proportion of the flagged detections came from observations which contained several detections flagged as containing QPEs. Of the 705 detections 484 ($69\%$ of all flagged detections) came from a combination of 37 observations, all of which contained at least 5 detections. Observation 0604960301 contained 46 detections which were flagged as containing QPEs at the conservative cuts implemented in the methodology described above. In the case of some of these observations there are unusual patterns affecting the detections which are flagged as containing QPEs. One example is that of observation 0804680101, where there are five detections flagged as containing QPEs which all show a similar profile with peaks simultaneously in their lightcurves. See Fig. \ref{fig:multi-det-obs-pattern} for the 250s binned lightcurves for sources 6, 7, 10, 11 and 22 from observation 0804680101. For this specific observation there are peaks across the lightcurves of all five objects simultaneously at $\sim15 \text{\,ks}$, $\sim24 \text{\,ks}$, $\sim75 \text{\,ks}$, $\sim100 \text{\,ks}$ and $\sim115 \text{\,ks}$ after the start of the observation. These features are evident in the lightcurves for all five sources, despite them being located on different parts of the detector, with source 6 being located on CCD 1, sources 7 and 10 being located on CCD 11, and sources 11 and 22 being located on CCD 10, and without overlapping raw positions on the detectors. The manner in which the lightcurves are created is standard across all detections in the catalogue. It requires the automatic detection of source locations, with background regions being drawn from empty regions on the same CCD as the source. Any other detections flagged in the background region are screened out. That sources located on different parts of the detector should have simultaneous, correlated variability is unlikely, and at present it is unknown as to what may be causing these features to appear across multiple detections in certain observations, but a manual recreation of the lightcurves for these detections appears to show that background features are not appropriately being subtracted.

\begin{figure}
	\includegraphics[width=\columnwidth]{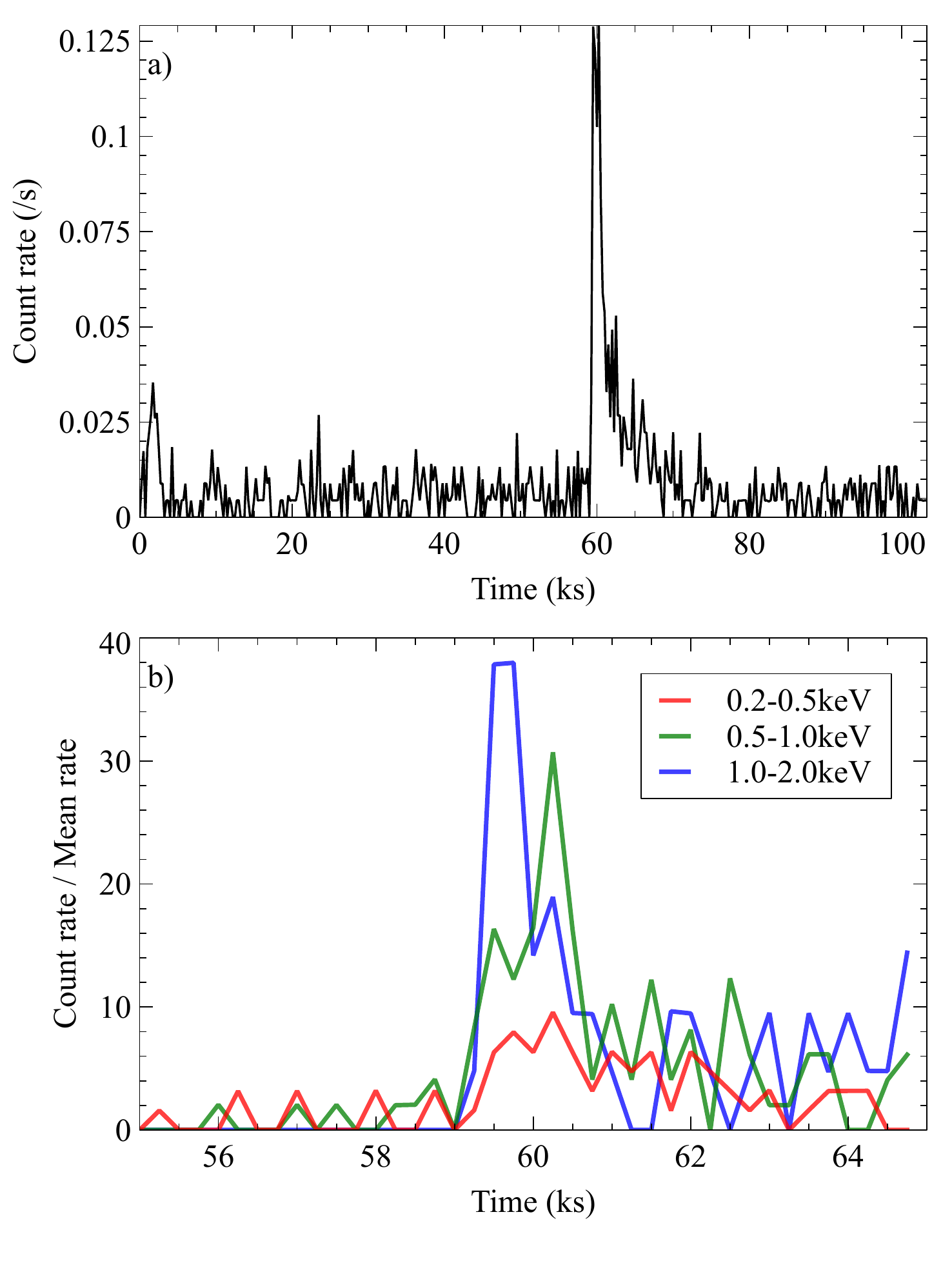}
    \caption{Lightcurves for the detection of 207223603010022 as source 22 in observation 0722360301. Top panel (a) shows the full band 0.2--12.0keV lightcurve binned at a rate of 50s for the source and the lower panel (b) shows the energy dependence of the flare located around 60ks after the start of the observation binned at a rate of 250s.}
    \label{fig:xmmssc-candflare}
\end{figure}

The 705 detections were then grouped by observation ID, to allow for faster manual filtering by repeated features across detections in the same observation, and were then all reviewed manually. From this sample we were able to identify 27 detections where their lightcurves warranted further examination. The details of these 27 detections are listed in Table \ref{tab:xmmssc-cands}. The detections were then cross-referenced with the SIMBAD astronomical database \citep{wenger_simbad_2000} to determine the names and types of sources to determine if the objects could be AGN as viable QPE sources. In the cases where there were no correlated source in SIMBAD we then referenced the identities of those sources through NED. Of these sources 19 were identifiable as stars or X-ray binary systems and not appropriate for further analysis to determine the presence of QPEs, and the remaining 8 were galaxies or unclassified sources. For those 8 sources we then downloaded and reprocessed the observation event lists, manually screened for flaring events, and created lightcurves for the 0.2--2.0keV  and 2.0--12.0keV bands to determine whether the variability was present in soft X-ray bands. Of those 8 sources the variability was not present at lower energies for detections 200823401010018, 204045402010025, and 206019305010009. For the remaining five sources we then created lightcurves in three narrower energy bands, 0.2--0.5keV, 0.5--1.0keV and 1.0--2.0keV, to determine whether the variability which appears in the soft X-rays follows the energy-dependent characteristics seen in confirmed QPE sources. At this stage we ruled out detections 200773401015192, 201068601010017, 206010101010020, 207277801010019, and 208235803010011 as there is no significant difference seen in the variability in the three energy bands, with the flare being the same width, peaking at the same time and having the same amplitude against the quiescent rate in all three narrower bands. We give an example of a source which has been flagged and examined in further detail in Fig. \ref{fig:xmmssc-candflare}. In this case the source identified (SRCID 207223603010022) by the automatic classification pipeline and manual inspection was classified as a red dwarf, likely displaying chromospheric flares, and as such no further analysis is required. At this stage our pipeline has not identified any new candidate QPE sources, with common sources of contamination being: instrumental background subtraction; chromospheric flares from red dwarfs; flares from other classes of variable stars.

\section{Conclusions}
\label{sec:conc}

In this analysis we set out to determine whether it would be possible to identify lightcurves containing quasi-periodic eruptions by means of time-domain only features and commonly used machine learning classification techniques. This analysis used the freely available \texttt{tensorflow} packages for \texttt{python} and features which had been previously used to classify other types of variability \citep{sokolovsky_comparative_2017,faisst_how_2019}, and were able to achieve accuracies of over 94$\%$ in classifying simulated lightcurves and accuracies of over 98$\%$ in classifying real observational data. The classification accuracy we have obtained is greater than that of \citet{lo_automatic_2014} ($\sim77\%$) when only considering time-domain features and also uses a smaller number of features (14 rather than 27). The accuracy we have achieved is even greater than that obtained by \citet{lo_automatic_2014} ($\sim97\%$) and \citet{farrell_autoclassification_2015} ($\sim92\%$) when considering non-temporal features, which we do not consider in this analysis, in addition to time-domain features. We chose not to use spectral features as part of the network training process for two reasons: simulating spectral features alongside temporal features would have been more computationally demanding; it could bias our classifier towards only finding QPE hosts with spectral features that are similar to those which have already been discovered. As all current QPE hosts have very soft spectra we could miss out on some harder X-ray sources of QPEs. As it is, the training on features sets takes of the order of minutes on a standard desktop computer, and the time taken for the creation of features and classification of lightcurves with the networks once trained is trivial.

\subsection{Further Work}
\label{subsec:further-work}

Further work could provide several ways to improve the efficacy of our classifier. Our approach to simulating lightcurves is itself limited by the number of eruptions which have currently been detected. With an increase in the number of known QPEs we could improve our understanding of the underlying distributions for the QPE features and so create a more accurate training data set. Our approach has also not required any information on the brightness of sources being considered, as this could be impacted by the sensitivity of different instruments. We should also consider the features being used in the classification process and whether the features created by the simulated data set adequately represent those seen in observational data. Feature 1 is likely to not be representative of real observational data for the reasons identified in section \ref{subsec:simlcfeatdists}. A further correction to the baseline of lightcurves after the QPE signals have been imposed could alter the effectiveness of and authenticity of this feature, but the issues caused by Poisson noise at very low count rates may mean that it would be better to eliminate the feature entirely. Features 13 and 14 also do not appear to follow the same distributions in the simulated and observational data sets, as shown in Fig. \ref{fig:dt50-featdists}, \ref{fig:dt250-featdists} and \ref{fig:dt1000-featdists}, and corrections to the generation process could be considered in order to make them more representative of the observational data. When we consider subsets of 13 of the 14 features we find that improved performance in classifying the observational data sample is possible. With 1000s binned lightcurves we can achieve an accuracy of $96.9\%$ when feature 3 is omitted. Further analysis of combinations of fewer features may identify the optimal approach moving forward. Additionally, we can consider data from instruments other than \emph{XMM Newton}. We have chosen not to include instrumental effects in generating lightcurves and features in order to allow this tool to be applied to data from other observatories, and including QPE features derived from observations of QPE sources with \emph{Chandra} and \emph{NICER} and other observatories may further improve the accuracy of these classifiers. An investigation into the effects of including Poisson noise in the populations of simulated lightcurves found a decrease in the ability of the classifiers to distinguish between lightcurves containing QPEs or not. The addition of noise increased the baseline variance of the lightcurves which made the two classes in training appear less distinct. However, at count rates typical of the datasets in which QPEs are detected, the degradation in performance was small. With the additional computational power and time which was required to include Poisson noise in simulating lightcurves we do not consider that it is likely to produce improvements in the classification accuracy in the future. Further applications of similar methods in other wavebands may allow for the detection of other types of astrophysical transient, and the network design with a SoftMax layer allows us to extend to such a multi-class problem. Additionally, it may be preferable when working with other catalogues of data to train the network using the $F_\beta$ metric. In this instance we found a slight decrease in the efficacy of the classifier when applied to real data when training with $F_\beta$, and the number of lightcurves flagged for manual inspection, when the networks were optimized for accuracy, in section \ref{subsec:xmmssc_res} was manageable. With a larger catalogue it may be beneficial to train with $F_\beta$ tuned to produce a smaller number of objects for manual inspection, even if the proportion of false positive results in that sample could be higher. 

\bigskip
Ultimately, this analysis can serve as a benchmark for the detection of quasi-periodic eruptions in X-ray lightcurves with machine learning. The classification has been achieved with high levels of accuracy in simulated lightcurve sets and high, albeit lower, accuracy in real observational data. While we were not able to conclusively identify any new QPE sources in the XMM Serendipitous Source Catalogue we have identified two sources which show some QPE-like behaviour which could be followed up with further observation.  

\bigskip
\section*{Acknowledgements}

All figures in this work were created with \texttt{Veusz} and \texttt{matplotlib}. This work is supported by the UKRI AIMLAC CDT, funded by grant EP/S023992/1. This research has made use of data obtained from the 4XMM XMM-Newton serendipitous source catalogue compiled by the XMM-Newton Survey Science Centre. This analysis used the specific version 4XMM-DR12. This research has made use of the SIMBAD database,
operated at CDS, Strasbourg, France. This research has made use of the NASA/IPAC Extragalactic Database (NED), which is funded by the National Aeronautics and Space Administration and operated by the California Institute of Technology.
The authors would like to thank the reviewers for their constructive comments during the review process. In particular we would like to thank Reviewer \#2 for their identification of two of the previously un-categorised sources flagged during the XMMSSC analysis.

%%%%%%%%%%%%%%%%%%%%%%%%%%%%%%%%%%%%%%%%%%%%%%%%%%
\section*{Data Availability}

All data used in this analysis is freely available in the \emph{XMM} Newton Science Archive (\url{http://nxsa.esac.esa.int/nxsa-web/}). All analysis tools and products are available at \url{https://github.com/robbie-webbe/ML_QPEs}.

For the purpose of open access, the author has applied a Creative Commons Attribution (CC BY) licence to any Author Accepted Manuscript version arising



\bibliographystyle{rasti}
\bibliography{ML_QPEs} % if your bibtex file is called example.bib

%\begin{thebibliography}{99}
%\end{thebibliography}

%%%%%%%%%%%%%%%%%%%%%%%%%%%%%%%%%%%%%%%%%%%%%%%%%%

%%%%%%%%%%%%%%%%% APPENDICES %%%%%%%%%%%%%%%%%%%%%

\appendix

\appendix

\section{Appendix A - Observations Used For Testing}
\label{appendix:A}

Table \ref{tab:all-obs1} contains a list of all \emph{XMM} observations of AGN which were processed and used in the real data testing phase. In all cases the pn lightcurves were binned at 50s, 250s and 1000s to account for shot-noise in lower count rate observations. Mass estimates are either taken from  or the CHANSNGCAT catalogue.

\begin{table}
	\centering
	\caption{Details of observations of low-mass AGN which were identified as part of a manual search for QPE candidates. Values for $M_\text{BH}$ are given as $\log(M_\text{BH} / M_{\sun})$ and are as per CHANSNGCAT. Those observations which could not be rebinned at 250s are denoted with (*), and those which could not be rebinned at 1000s are denoted with ($\dagger$) against their observation ID numbers.}
	\label{tab:all-obs1}
	\begin{tabular}{cclr} % 
		\hline
		AGN Name & $M_\text{BH}$ & OBSID & Exp. (ks) \\
		\hline
            2XMM J123103.2+110648 & 4.87 & 0145800101 & 107.0\\
            -- & -- & 0306630101 & 80.9\\
            -- & -- & 0306630201 & 99.5\\
            eRASSU J023147.2-102010 & 5.78 & 0861910201 & 94.2\\
            -- & -- & 0861910301 & 90.2\\
            eRASSU J023448.9-441931 & 4.96 & 0872390101 & 95.0\\
            -- & -- & 0893810501 & 25.0\\
            GSN 069 & 5.99 & 0657820101 & 14.9\\
            -- & -- & 0740960101 & 95.1\\
            -- & -- & 0823680101 & 63.3\\
		-- & -- & 0831790701 & 141.4\\
            -- & -- & 0851180401 & 135.4\\
            -- & -- & 0864330101 & 141.0\\
            -- & -- & 0864330201 & 133.1\\
            -- & -- & 0864330301 & 133.2\\
            -- & -- & 0864330401 & 136.1\\
		NGC1331 & 5.48 & 0304190101 & 65.9 \\
		NGC3367 & 5.64 & 0551450101 & 31.4 \\
		NGC3599 & 5.85 & 0411980101 & 6.9 \\
		 -- & --  & 0556090101 & 43.6 \\
		NGC4467 & 5.86 & 0112550601 & 24.6 \\
		 -- & -- & 0510011501 ($\dagger$) & 10.1 \\
		 -- & -- & 0761630101 & 118.0 \\
		 -- & -- & 0761630201 & 118.0 \\
		 -- & -- & 0761630301 & 117.0 \\
		\hline
	\end{tabular}
\end{table}

\begin{table}
	\centering
	\contcaption{: Details of observations of low-mass AGN which were identified as part of a manual search for QPE candidates. Values for $M_\text{BH}$ are given as $\log(M_\text{BH} / M_{\sun})$ and are as per CHANSNGCAT. Those observations which could not be rebinned at 250s are denoted with (*), and those which could not be rebinned at 1000s are denoted with ($\dagger$) against their observation ID numbers.}
	\label{tab:all-obs2}
	\begin{tabular}{cclr} % 
		\hline
		AGN Name & $M_\text{BH}$ & OBSID & Exp. (ks) \\
		\hline
		NGC4476 & 5.58 & 0200920101 & 109.3 \\
		 -- & -- & 0551870401 & 21.6 \\
		 -- & -- & 0551870601 & 21.3 \\
		 -- & -- & 0603260201 & 17.9 \\
		 -- & -- & 0803670501 & 132.0 \\
		 -- & -- & 0803670601 & 65.0 \\
		 -- & -- & 0803671001 & 63.0 \\
		 -- & -- & 0803671101 & 131.9 \\
		NGC4559 & 5.14 & 0152170501 & 42.2 \\
		 -- & -- & 0842340201 & 75.4 \\
		NGC4654 & 5.07 & 0651790201 & 28.9 \\
		NGC5273 & 5.97 & 0112551701 & 17.1 \\
		 -- & -- & 0805080401 & 110.9 \\
		 -- & -- & 0805080501 & 28.0 \\
		NGC6946 & 5.43 & 0093641501 (*) & 8.6 \\
		 -- & -- & 0093641601 (*) & 10.1 \\
		 -- & -- & 0093641701 (*) & 11.3 \\
		 -- & -- & 0200670101 & 16.4 \\
		 -- & -- & 0200670201 ($\dagger$) & 14.4 \\
		 -- & -- & 0200670301 & 15.6 \\
		 -- & -- & 0200670401 & 21.2 \\
		 -- & -- & 0401360101 & 20.9 \\
		 -- & -- & 0401360201 & 24.4 \\
		 -- & -- & 0401360301 & 24.4 \\
		 -- & -- & 0500730101 & 31.9 \\
		 -- & -- & 0500730201 & 37.3 \\
		 -- & -- & 0691570101 & 119.3 \\
		 -- & -- & 0794581201 & 50.0 \\
		 -- & -- & 0870830101 & 17.9 \\
		 -- & -- & 0870830201 & 17.7 \\
		 -- & -- & 0870830301 & 16.0 \\
		 -- & -- & 0870830401 & 17.8 \\
		NGC7314 & 5.59 & 0111790101 & 44.7 \\
		 -- & -- & 0311190101 & 83.9 \\
		 -- & -- & 0725200101 & 140.5 \\
		 -- & -- & 0725200301 & 132.1 \\
		 -- & -- & 0790650101 & 65.0 \\
		NGC925 & 6.0 & 0784510301 & 50.0 \\
		 -- & -- & 0862760201 & 42.0 \\
            RX J1301.9+2747 & 6.65 & 0124710801 & 29.8\\
            -- & -- & 0851180501 & 48.4\\
            -- & -- & 0864560101 & 134.9\\
            XMMSL1 J024916.6-041244 & 5.29 & 0411980401 & 11.7\\
            -- & -- & 0891800601 & 33.8\\
		\hline
	\end{tabular}
\end{table}

\section{Appendix B - Top Candidates from XMM SSC Analysis}
\label{appendix:B}

Table \ref{tab:xmmssc-cands} contains details of those detections from the XMM Serendipitous Source Catalogue (4XMM-DR12) which were manually identified as containing significant variability and where follow-up analysis was performed. Table \ref{tab:xmmssc-cands} lists the type of object corresponding to the detection and whether it contains quasi-periodic eruptions if a galaxy.

\begin{table*}
	\centering
	\caption{Details of detections from the XMM SSC which were flagged as containing QPEs and were followed up with manual inspection. The last three columns contain details as to the type of astrophysical object, whether the variability is seen when only the 0.2--2.0keV band lightcurve is created, and whether the energy dependent and temporal characteristics are indicative of QPE behaviour. Sources are listed in order of their 4XMM Catalogue source IDs.}
	\label{tab:xmmssc-cands}
	\begin{tabular}{ccccccc} % 
		\hline
		SRCID & OBSID & Source No. & Object Name & Type & Soft Variability & QPE-like? \\
		\hline
	    200287402010079 & 0655050101 & 11 & 2XMM J001527.9-390507 & Star & N/A & N/A \\
           200560203010005 & 0801610101 & 4 & [BHR2005] 832-14 & Low-Mass Star & N/A  & N/A \\
           200669401010002 & 0501790101 & 2 & Cl* NGC 2547 JND 13-98 & Star & N/A  & N/A \\
           200669401010075 & 0501790101 & 15 & Cl* NGC 2547 NTJ 7-2323 & Low-Mass Star & N/A  & N/A \\
           200773401015192 & 0555630401 & 136 & 2XMM J150122.1-414227 & X-ray Source & Y & N \\
           200823401010018 & 0203560201 & 12 & SDSS J111740.11+074411.7 & Quasar & N & N/A \\
           201068601010017 & 0803990301 & 4 & [CPS95] X-12 & X-ray Source & Y & N \\
           201080604010030 & 0555780601 & 5 & [LBX2017] 780 & High Proper Motion Star & N/A  & N/A \\
           -- & 0604960801 & 3 & -- & -- & -- & -- \\
           -- & 0604961201 & 5 & -- & -- & -- & -- \\
           201111202010026 & 0305540701 & 14 & [GY92] 463 & T-Tauri Star & N/A  & N/A \\
           201111202010048 & 0800031001 & 3 & [GY92] 259 & Young Stellar Object & N/A  & N/A \\
           201125903010151 & 0403200101 & 85 & V* V1320 Ori & BY Dra Variable & N/A  & N/A \\
           201428001010001 & 0142800101 & 1 & 
X LMC X-4 & High-Mass XRB & N/A & N/A \\
           202010902010004 & 0821240301 & 3 & WISEA J020621.12-002346.7 & Star & N/A  & N/A \\
           204045402010025 & 0674050101 & 34 & 2CXO J202103.2+365423 & X-ray Source & N & N/A \\
           206006901010002 & 0600690101 & 2 & UCAC4 509-131194 & High Proper Motion Star & N/A  & N/A \\
           206010101010020 & 0844860101 & 8 & XMMU J004705.9-205239 & X-ray Source & Y & N \\
           206019305010009 & 0601930501 & 6 & WISEA J213740.24+002048.0 & IR Source & N & N/A \\
           206048602010005 & 0604860201 & 5 & WISEA J181323.39-325230.9 & Red Dwarf & N/A & N/A \\
           206939901010022 & 0693990301 & 10 & WISEA J111949.13+065305.6 & Star & N/A & N/A \\
           207216201010018 & 0721620101 & 18 & 2MASS J08384128+1959471 & Eruptive Variable Star & N/A  & N/A \\
           207223603010022 & 0722360301 & 22 & WISEA J220310.58-344406.7 & Red Dwarf & N/A & N/A \\
           207277801010019 & 0790380901 & 18 & [ELK2021] 22 & Galaxy & Y & N \\
           207437002010001 & 0743700201 & 1 & TYC 4682-1697-1 & Star & N/A & N/A \\
           207810401010007 & 0781040101 & 7 & SDSS J032048.68+003234.0 & Low-Mass Star & N/A  & N/A \\
           208235803010011 & 0823580301 & 11 & WISEA J124901.51-410131.6 & IR Source & Y & N \\
		\hline
	\end{tabular}
\end{table*}

%%%%%%%%%%%%%%%%%%%%%%%%%%%%%%%%%%%%%%%%%%%%%%%%%%

% Don't change these lines
\bsp	% typesetting comment
\label{lastpage}
\end{document}